\documentclass[11pt]{article}
\usepackage{jheppub,amsmath, amsthm, amssymb,slashed,url}
\usepackage{graphicx}
\usepackage{epstopdf}

\usepackage{kbordermatrix}% http://www.hss.caltech.edu/~kcb/LaTeX.shtml

\def\barre#1{{\not\mathrel #1}}

\def\qs{\barre {{\!}q}}

\def\OMIT#1{{}}

\newcommand{\bit}[1]{\mbox{\boldmath$#1$}}
\newcommand{\ft}[2]{{\textstyle\frac{#1}{#2}}}

\def\OMIT#1{{}}

\def\yo2{{f_\pi^2}}
\def\llra{{\relbar\joinrel\longrightarrow}}
\def\mapright#1{{\smash{\mathop{\llra}\limits_{#1}}}}
\def\mapwrong#1{{\smash{\mathop{\;\;=}\limits_{#1}}}}

\def\oneht{\textstyle{1\over 2} }
\def\oneft{\textstyle{1\over 4} }

\def\oneft{\textstyle{1\over 4} }

\def\vev#1{{\langle #1\rangle}}

\newcommand{\BE}{\begin{equation}}
\newcommand{\EE}{\end{equation}}
\newcommand{\BA}{\begin{eqnarray}}
\newcommand{\EA}{\end{eqnarray}}

\font\teneurm=eurm10 \font\seveneurm=eurm7 \font\fiveeurm=eurm5
\newfam\eurmfam
\textfont\eurmfam=\teneurm \scriptfont\eurmfam=\seveneurm
\scriptscriptfont\eurmfam=\fiveeurm

\font\teneusm=eusm10 \font\seveneusm=eusm7 \font\fiveeusm=eusm5
\newfam\eusmfam
\textfont\eusmfam=\teneusm \scriptfont\eusmfam=\seveneusm
\scriptscriptfont\eusmfam=\fiveeusm

\font\tencmmib=cmmib10 \skewchar\tencmmib='177
\font\sevencmmib=cmmib7 \skewchar\sevencmmib='177
\font\fivecmmib=cmmib5 \skewchar\fivecmmib='177
\newfam\cmmibfam
\textfont\cmmibfam=\tencmmib \scriptfont\cmmibfam=\sevencmmib
\scriptscriptfont\cmmibfam=\fivecmmib

\def\Pi{\varPi}
\dedicated{NT@UW-15-15}

\title{Aspects of QCD Current Algebra on a Null Plane}

 \author{S.R.~Beane and T.J.~Hobbs}

\affiliation{Department of Physics, University of Washington,
Seattle, WA 98195}

\date{\mydate}

\abstract{Consequences of QCD current algebra formulated on a
  light-like hyperplane are derived for the forward scattering of
  vector and axial-vector currents on an arbitrary hadronic target.
  It is shown that current algebra gives rise to a special class of
  sum rules that are direct consequences of the independent chiral
  symmetry that exists at every point on the two-dimensional
  transverse plane orthogonal to the lightlike direction. These sum
  rules are obtained by exploiting the closed, infinite-dimensional
  algebra satisfied by the transverse moments of null-plane
  axial-vector and vector charge distributions.  In the special case
  of a nucleon target, this procedure leads to the Adler-Weisberger,
  Gerasimov-Drell-Hearn, Cabbibo-Radicatti and Fubini-Furlan-Rossetti
  sum rules.  Matching to the dispersion-theoretic language which is
  usually invoked in deriving these sum rules, the moment sum rules
  are shown to be equivalent to algebraic constraints on forward
  S-matrix elements in the Regge limit.}

\begin{document} \maketitle

%\vfill\eject
%%%%%%%%%%%%%%%%%%%%%%%%%%%%%%%%%%%%%%%%%%%%%%%%%%%%%%%%%%%%%%%%%%%%%%%%%%%%%%%%%%
\section{Introduction}
\label{intro}

\noindent Consequences of the chiral symmetries of QCD for nucleon
structure are usually viewed impressionistically as arising from the
presence of a ``pion cloud'' that surrounds the nucleon, whose
quantitative physical effects can be calculated systematically using
chiral perturbation theory ($\chi$PT), an expansion in small momenta
and quark masses~\cite{Weinberg:1978kz,Gasser:1983yg}. However, there
are other consequences of QCD chiral symmetries that are difficult to
access using effective field theory, as the chiral constraints occur
in the form of sum rules which correlate many hadronic length
scales~\cite{Weinberg:1969hw,Weinberg:1969db,Weinberg:1990xn}.  These
sum rules are profitably studied by considering matrix elements of the
QCD chiral symmetry currents and in particular of the Lie brackets
that these currents obey on the light-cone. This paper provides
a unified and self-contained derivation of a special class of
chiral sum rules which constrain the scattering of currents on
arbitrary hadronic targets in a manner consistent with QCD and its
symmetries. 

The sum rules that are obtained distinguish themselves from other
current-algebra sum rules in that they are entirely consequences of
symmetry. Many of the relations that we derive are known, however
their modern interpretation is somewhat murky in the sense that the
connection to the symmetries of the underlying theory is obscured.
Consider, for example, the Gerasimov-Drell-Hearn (GDH) sum
rule~\cite{Gerasimov:1965et,Drell:1966jv} and the Adler-Weisberger
(AW) sum rule~\cite{Adler:1965ka,Weisberger:1965hp} for the
nucleon. In the modern viewpoint, these sum rules are obtained as
follows. One derives soft-photon and soft-pion theorems (the latter
using $\chi$PT), respectively, to determine a scattering amplitude at
low incident photon or pion energy. The machinery of dispersion
relations then relates the scattering amplitude at the special
low-energy kinematical point to the integral of the imaginary part of
the amplitude over all energies. The caveat here is that the
convergence of the integral requires one to assume an asymptotic
behavior of the scattering amplitude that is softer than that required
by unitarity via the Froissart-Martin
bound~\cite{Froissart:1961ux,Martin:1962rt}. As these sum rules are
for forward scattering, the asymptotic behavior is governed by the
Regge limit ($s\gg -t$) and therefore general assumptions that are
suggested by models like Regge pole theory are invoked to motivate the
desired asymptotic behavior. While this logic is unassailable, this
point of view obscures the origin of the sum rule by introducing
mysterious model dependence which is actually not present.  Indeed, it
will be shown in this paper that the asymptotic behavior of a class of
amplitudes which scatter the QCD vector and axial-vector currents on
arbitrary hadronic targets in the Regge limit is a direct consequence
of the QCD algebra of currents. In practice, this implies that these
scattering amplitudes in the Regge limit can be expressed in terms of
null-plane QCD operator structures that vanish in the chiral symmetry
limit. 

A useful tool for establishing the chiral sum rules is null-plane
quantization (front-form
dynamics)~\cite{Dirac:1949cp,Kogut:1969xa,Leutwyler:1977vy}.  For reasons
that will be made clear in the text, in
null-plane coordinates it is possible to give a transparent and
frame-independent derivation of the sum rules\footnote{ The same
  results can be obtained in the usual equal-time quantization
  (instant-form dynamics)~\cite{Dirac:1949cp}. However, this requires one to work in
  special (fictitious) Lorentz frames in which the system is boosted to
  infinite momentum~\cite{deAlfaro:1973zz}.  In addition to the lack
  of frame independence, there are subtle issues in taking the limit
  which sometimes lead to incorrect results~\cite{Dicus:1971uk}.}.
While the derivation of the sum rules is independent of considerations
of the hadronic constituents, it is well known that the null plane formulation lends itself
to an intuitive picture of a hadron in terms of partonic degrees of freedom\footnote{See, for instance, Ref.~\cite{Burkardt:2002hr,Diehl:2003ny,Belitsky:2005qn}.}.

It is useful to build some physical intuition to understand why chiral
and gauge symmetries should lead to sum rules, and, in particular, to
understand why null planes, which are quantization surfaces designed
for the study of high-energy processes, are useful for obtaining these
sum rules. As an example, consider the low-energy scattering of a pion
with an arbitrary hadronic target. In QCD this process is described in
a model-independent manner using $\chi$PT. Beyond leading order in
$\chi$PT, there arise operators whose couplings appear not to be
constrained by chiral symmetry, and which must therefore be fixed by
experiment or through lattice QCD simulations. As the momentum of the
external pion gets larger there is some point at which the
chiral expansion fails and predictive power is lost. Now if one
chooses kinematics with fixed-$t$ and large-$s$, then at large enough
$s$, one expects that the scattering amplitude will behave as dictated
by Regge lore. Therefore, as one continuously varies the momentum from
low to high, the scattering amplitudes must change from polynomial
growth in momentum, to soft behavior that is at least as soft as
required by unitarity and sometimes softer. That this occurs is not
entirely surprising as new degrees of freedom must be integrated in as
momentum increases beyond new production thresholds. Nevertheless,
these new states must have masses and couplings that are related to
those of the pions and the hadronic target in order that
the necessary cancellations take place which give rise to Regge
behavior in the asymptotic
limit~\cite{Weinberg:1969hw,Weinberg:1969db,Weinberg:1990xn,Weinberg:1994tu}.
The procedure outlined above was carried out explicitly by Weinberg,
who wrote down the most general chiral Lagrangian which gives rise to
the scattering of pions on arbitrary hadronic targets in
the forward limit, calculated the scattering amplitude involving the
sum of all tree graphs and, extracted --via contour integration-- the
asymptotic behavior of the amplitude and found that it takes a
manifestly Lie algebraic structure consistent with the full chiral
symmetry group. Since the Lie algebraic form of the constraint appears
as the coefficient of the leading term in a high-energy expansion, it
is not particularly surprising that null-plane quantization becomes a
useful tool to derive the sum rules directly from the QCD current algebra.
Indeed, it will be shown that Weinberg's results emerge as a special
case when the sum rules are saturated with single-particle states.

The argument given above may seem inconsistent to the $\chi$PT
practitioner since the low-energy constants of $\chi$PT are, naively,
not constrained by chiral symmetry. That is, $\chi$PT is constructed
by choosing definitions of fields such that all of the chiral symmetry
resides in the pion fields and all other fields transform according to
the unbroken vector subgroup. As emphasized by Weinberg, this seeming
contradiction arises because symmetries like chiral symmetry (and
gauge invariance and general coordinate invariance) have two
manifestations which lead to distinct constraints on physical
observables.  Weinberg has referred to these two kinds of symmetry
constraints as ``dynamical'' and ``algebraic'',
respectively~\cite{Weinberg:1969hw,Weinberg:1971}. A dynamical
symmetry is a symmetry of the Lagrangian which is not a symmetry of
the physical Hilbert space and manifests itself through low-energy
theorems.  An algebraic symmetry is a physical symmetry of Hilbert
space and manifests itself through algebraic constraints on the
S-matrix. It is interesting that one of these manifestations appears
naturally in the instant-form of dynamics, while the other appears
naturally in the front-form.  Specifically, instant-form chiral
symmetry constrains low-energy processes with pions on external lines
via $\chi$PT, whereas front-form chiral symmetry constrains the
asymptotic behavior of a class of scattering amplitudes with pions on
external lines. Indeed, there is an even more general kind of
complementarity between the two descriptions.  In the instant form,
which is conveniently formulated in terms of Lorentz invariant, local
quantum field theory, the dynamical consequences of chiral symmetry
are manifest, which, for instance, give rise to the consequences of
chiral symmetry that are encoded in $\chi$PT via Lagrangian field
theory, while the algebraic consequences of chiral symmetry are
enigmatic and therefore are usually assumed or modelled (for instance,
using Regge pole theory). By contrast, in the front form, the
situation is reversed; chiral symmetry constrains the asymptotic
behavior of the S-matrix while Lorentz invariance and locality (in the
longitudinal coordinate) are hidden in the same sense that the
hadronic spectrum is hidden, and the chiral-symmetry transformation
properties of the quark mass matrix, which lead to $\chi$PT, are
obscured. While physics is clearly independent of the choice of
coordinates, one may suspect that in considering a theory like QCD,
whose solution is unknown, the use of distinct quantization surfaces
would prove fruitful. This paper provides evidence that this is indeed
the case.

The null-plane description is tailored to describe hadrons or their
constituent partons that are moving very fast (large longitudinal
momentum $p^+$) relative to the vacuum ($p^+=0$).  It is then
difficult to visualize how spontaneous breaking of chiral symmetry can
occur on null-planes and communicate itself to the hadron and
constrain its
structure~\cite{Casher:1974xd,Casher:1973vh,Susskind:1994wr}.  One of
the drawbacks of the null-plane description has been the lack of a
consistent formalism to think about chiral symmetry breaking. Building
on earlier
work~\cite{Jersak:1969zg,Eichten:1973ip,Feinberg:1973qb,Carlitz:1974sg,Sazdjian:1974gk,Kim:1994rm,Yamawaki:1998cy,Wu:2003vn},
recent studies~\cite{Beane:2013oia,Beane:2015uaa} have considered
fundamental issues of chiral symmetry breaking in detail in a
model-independent manner, including a proof of Goldstone's theorem and
the issue of chiral condensates and how they appear on the null
plane. In some sense, what follows is a continuation of this previous
work with an emphasis on the observable consequences of chiral
symmetry of the algebraic type which arise from considerations of the
full current algebra.

This paper is organized as follows. Section~\ref{NPQCDconB}
establishes the necessary null-plane QCD notation.  The QCD current
algebra that gives rise to the chiral sum rules and its non-canonical
modifications is obtained in Section~\ref{NPQCDconC}, and current
algebra is reduced to algebraic constraints on the moments of vector
and axial-vector charge distributions on the transverse plane in
Section~\ref{NPQCDconC2}. The basic kinematics and technology
necessary for the sum rule derivations are set up in
Section~\ref{fme}.  The simplest moment sum rules are obtained in
Section~\ref{msr} for Lie brackets of axial-axial (Section~\ref{aa}),
vector-vector (Section~\ref{vv}) and axial-vector (Section~\ref{av}
and Section~\ref{av2}) type, both in the general case of arbitrary
hadronic targets, and in the special case of a nucleon target.
Section~\ref{sec:Ssymms} re-expresses the sum rules in terms of the
asymptotic behavior of the relevant forward scattering amplitudes in
the Regge limit. This demonstrates that the asymptotic behavior
derived from null-plane QCD is consistent with Regge lore and can be
understood as arising directly from symmetry constraints on the S-matrix.  In
Section~\ref{sec:concs} the conclusions are summarized and
discussed. Much of the technology and notation is taken from
Ref.~\cite{Beane:2013oia}. Basic null-plane coordinate conventions,
and our conventions for the various form-factors and scattering
amplitudes are relegated to appendices.

The reader who is familiar with history will find much that is
familiar in this paper. However, while many of the results that are
derived are known in the literature, they are quite distinct in
origin, both as regards the technology with which they are derived and
also as regards the manner in which they are applied. One of the goals
of this paper is to provide a unifying framework for the derivation of
the sum rules whose content can be directly tracked to QCD symmetries.

%%%%%%%%%%%%%%%%%%%%%%%%%%%%%%%%%%%%%%%%%%%%%%%%%%%%%%%%%%%%%%%%%%%%%%%%%%%%%%%%%%
\section{Null-plane QCD constraints}
\label{NPQCDcon}

\subsection{Chiral symmetry and currents}
\label{NPQCDconB}

\noindent Consider QCD with two degenerate flavors of light quarks. In the chiral limit of massless quarks this theory
has an ${SU}(2)_L\otimes {SU}(2)_R$ invariance
\begin{equation}
\psi(x)\rightarrow e^{-i\theta_\alpha T_\alpha} \psi(x) \ \ \ , \ \ \ \psi(x)\rightarrow e^{-i\theta_\alpha T_\alpha \gamma_5} \psi(x) \ ,
\end{equation}
where $\psi(x)$ is the isodoublet quark field, and $T_\alpha =\tau_\alpha/2$. The corresponding instant-form Noether currents are
\begin{equation}
{J}^\mu_\alpha(x)\ =\ \bar\psi(x) \gamma^\mu T_\alpha \psi(x) \ \ \ , \ \ \ 
{J}^\mu_{5\alpha}(x)\ =\ \bar\psi(x) \gamma^\mu\gamma_5 T_\alpha \psi(x) \ .
\end{equation}
Both of these currents are conserved in the chiral limit.

In null-plane quantization (see Appendix~\ref{NPQCDconA} for
coordinate conventions) the non-dynamical degrees of freedom are
integrated out leaving behind the dynamical gluon field and the dynamical quark fields,
$\psi^+\equiv {\mit\Pi}^+\psi$, where the projection operator is
defined as ${\mit\Pi}^{\pm} \equiv \ft12 \gamma^{\mp} \gamma^{\pm}$
and $\gamma^+ \equiv \gamma \cdot n$ and $\gamma^- \equiv \gamma \cdot
{\bar n}$.  At equal null-plane time, the dynamical quark field satisfies
\begin{eqnarray}
\{ \psi_+(x)\, ,\, \psi^\dagger_+(y)\}|_{x^+=y^+}
\ = \
\ft{1}{\sqrt{2}} {\mit\Pi}^+\;
\delta (x^- - y^-)
\delta^{2} (\bit{x}_\perp - \bit{y}_\perp) \ .
\end{eqnarray}

The presence of the ${SU}(2)_L\otimes {SU}(2)_R$ invariance is of
course independent of the choice of initial quantization surface and
indeed in null-plane QCD the chiral transformations are
\begin{equation}
\psi_+(x)\rightarrow e^{-i\theta_\alpha T_\alpha} \psi_+(x) \ \ \ , \ \ \ \psi_+(x)\rightarrow e^{-i\theta_\alpha T_\alpha \gamma_5} \psi_+(x) \ ,
\label{chiraltrans}
\end{equation}
which give rise to the front-form (tilded) Noether currents ${\tilde J}^\mu_{(5)\alpha}(x)$. 
While the vector currents are independent of the choice of coordinates, ${\tilde J}^\mu_{\alpha}(x)=J^\mu_{\alpha}(x)$,
the instant-form and front-form axial currents only share the $+$ component, ${\tilde J}^+_{5\alpha}(x) = {J}^+_{5\alpha}(x)$,
a reflection of the fundamentally important property that ${\tilde J}^\mu_{5\alpha}(x)$ is not conserved,
\BA
\partial_\mu {\tilde J}^\mu_{5\alpha}(x)\ \equiv \ {\tilde {\cal D}}_{5\alpha}(x) \ =\ F_\pi M_\pi^2 {\tilde\pi}_\alpha(x) \ ,
\label{eq:chidivqcd}
\EA 
even in the chiral
limit~\cite{Wu:2003vn,Beane:2013oia,Beane:2015uaa}. It is important to
stress that in null-plane quantization, the chiral limit should be taken
only {\it after} taking matrix elements of the pion interpolating
operator ${\tilde\pi}_\alpha(x)$. Using LSZ reduction it is easy to
see that these matrix elements scale as
$M_\pi^{-2}$~\cite{Beane:2013oia}, which immediately implies that the
matrix element of the axial-vector current divergence is non-vanishing
and independent of $M_\pi$.

In what follows only the null-plane vector and axial-vector charge distributions,
\begin{equation}
{\tilde J}^\mu_\alpha(x)\ =\ \bar\psi_+(x) \gamma^\mu T_\alpha \psi_+(x)\ =\  {J}^\mu_\alpha(x)\ \ \ , \ \ \ 
{\tilde J}^+_{5\alpha}(x)\ =\ \bar\psi_+(x) \gamma^+\gamma_5 T_\alpha \psi_+(x) \ =\ {J}^+_{5\alpha}(x) \ ,
\end{equation}
will be considered. For a complete catalog of notation, the reader should consult Ref.~\cite{Beane:2013oia}. 

\subsection{Current algebra on the light-cone}
\label{NPQCDconC}

\begin{figure}[!t]
  \centering
     \includegraphics[scale=0.31]{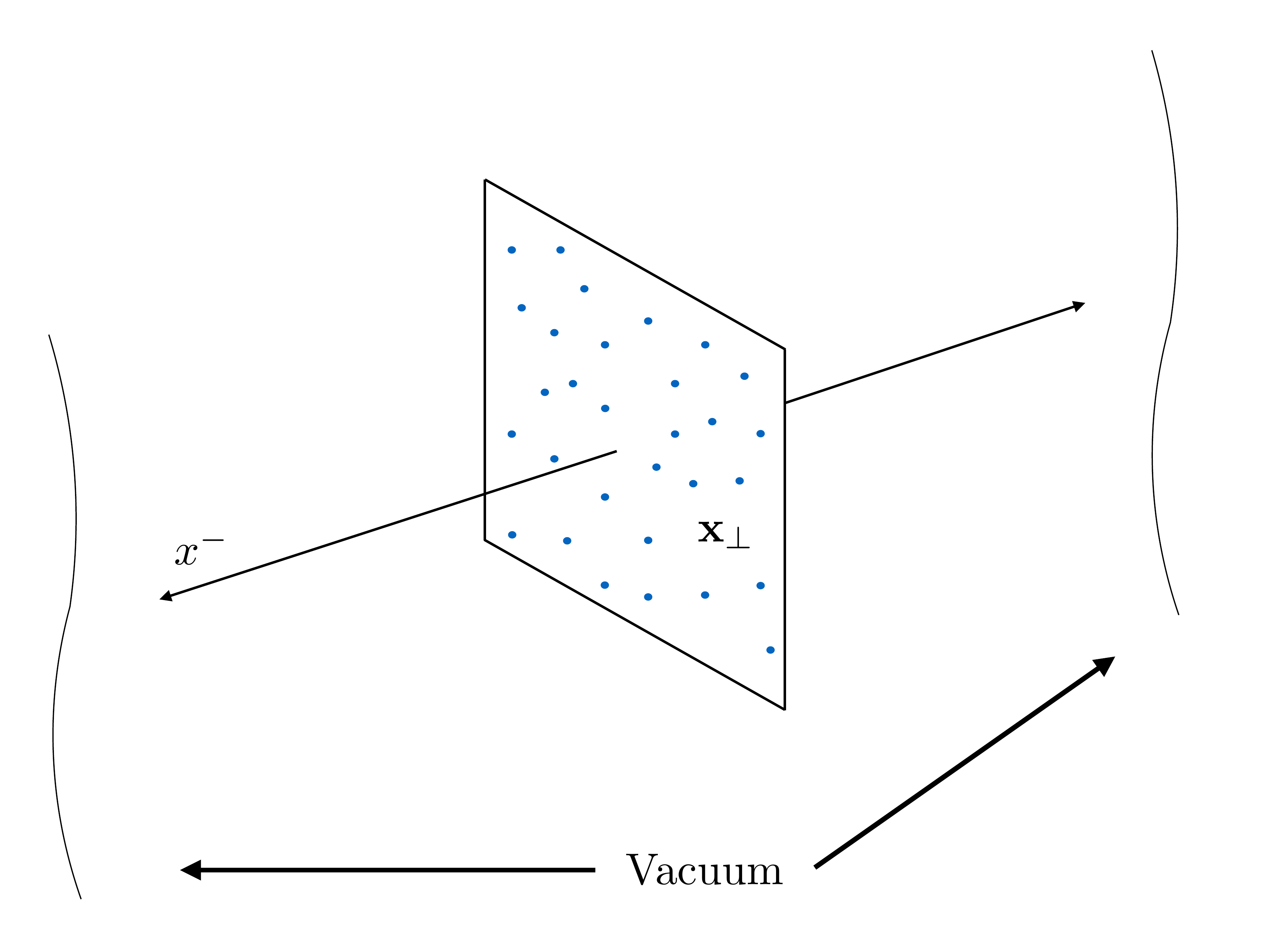}
     \caption{The dots represent the projections of partons, or sources of vector and axial charge, onto the transverse plane, specified by the coordinates ${\bf x_\perp}=(x^1,x^2)$. The vacuum physics is at the boundary
of the longitudinal coordinate, $x^-$.}
  \label{fig:transplane}
\end{figure}

\noindent Following the canonical procedure, the null-plane time components of the vector and axial charge distributions
are found to satisfy commutation relations at equal null-plane time\footnote{Specifying the commutator at equal null-plane time is
equivalent to specifying it on the light cone since with $x^+=y^+$, $(x-y)^2=-({\bf x}_\perp-{\bf y}_\perp)^2<0$ and therefore the commutator
vanishes by causality unless $(x-y)^2=0$.}:
\BA
{[\, {\tilde J}^+_{\alpha}(x)\,  ,\, {\tilde J}^+_{\beta}(y)\, ]|_{x^+=y^+}}\, &=& \, i\,\epsilon_{\alpha\beta\gamma} \, {\tilde J}^+_{\gamma}(x)
\delta(x^- - y^-)\,\delta^2(\bit{x}_\perp - \bit{y}_\perp )\ ;
\label{eq:LCalga}\\
{[\, {\tilde J}^+_{5\alpha}(x)\,  ,\, {\tilde J}^+_{\beta}(y)\, ]|_{x^+=y^+}}\, &=& \, i\,\epsilon_{\alpha\beta\gamma} \, {\tilde J}^+_{5\gamma}(x)
\delta(x^- - y^-)\,\delta^2(\bit{x}_\perp - \bit{y}_\perp)\ ;
\label{eq:LCalgb}\\
{[\, {\tilde J}^+_{5\alpha}(x)\,  ,\, {\tilde J}^+_{5\beta}(y)\, ]|_{x^+=y^+}}\, &=& \, i\,\epsilon_{\alpha\beta\gamma} \, {\tilde J}^+_{\gamma}(x)
\delta(x^- - y^-)\,\delta^2(\bit{x}_\perp - \bit{y}_\perp)\ .
\label{eq:LCalgc}
\EA As the charge distributions satisfy the chiral algebra at every
point in space\footnote{As the structure constants are highly
  singular, one can interpret this statement in the regulated sense
  where space is replaced by a set of discrete points and the
  commutators become, for instance, $[\, {\tilde J}^+_{\alpha}(i)\,
  ,\, {\tilde J}^+_{\beta}(j)\, ]= i\,\epsilon_{\alpha\beta\gamma} \,
  {\tilde J}^+_{\gamma}(j)\delta_{ij}$.}, the associated symmetry is
infinite dimensional.  However, it is clear that there are important
terms missing from the canonical procedure which must be present on
general grounds~\cite{Cornwall:1971as,Dicus:1971uk,Jackiw:1972ee}.
The vacuum expectation values of the commutators of the currents have
the well-known spectral decomposition~\cite{Weinberg:1996kr}
\BA
&&{\hspace{-0.45in}}{\langle\, 0\, | {[\, {J}^\mu_{\alpha}(x)\,  ,\, {J}^\nu_{\beta}(0)\, ]}|\, 0\,\rangle} =
\delta_{\alpha\beta}\!\!\!\int d\lambda^2 \left( g^{\mu\nu} - {\partial^\mu \partial^\nu}/{\lambda^2} \right) \rho^{(1)}_{V}(\lambda^2)\Delta(x;\lambda^2) \ ; \label{eq:LCalgaKLa} \\
&&{\hspace{-0.45in}}{\langle\, 0\, | {[\, {J}^\mu_{5\alpha}(x)\,  ,\, {J}^\nu_{5\beta}(0)\, ]}|\, 0\,\rangle} =
\delta_{\alpha\beta}\!\!\!\int d\lambda^2 \Big\lbrack\left( g^{\mu\nu} - {\partial^\mu \partial^\nu}/{\lambda^2} \right) \rho^{(1)}_{A}(\lambda^2) - {\partial^\mu \partial^\nu}\rho^{(0)}_{A}(\lambda^2)\Big\rbrack\Delta(x;\lambda^2) 
\label{eq:LCalgaKLb}
\EA
where $\rho^{(S)}_{V,A}$ are the spin-$S$ spectral functions, and the free-field commutator function is
\BA
\Delta(x;\lambda^2) \ =\ \frac{1}{(2\pi)^3}\int d^4k\;\varepsilon (k^0)\delta(k^2-\lambda^2)e^{-ikx} \ \mapwrong{x^+=0}\ -\frac{i}{4}\varepsilon( x^-)\delta^2({\bit{x}_\perp}) \ .
\label{eq:FFCF}
\EA
where $\varepsilon( z)$ is the sign function.
The vacuum expectation values in Eqs.~(\ref{eq:LCalgaKLa}) and (\ref{eq:LCalgaKLb}) cannot vanish due to positivity
constraints. However, the vacuum expectation values of the right-hand sides of Eqs.~(\ref{eq:LCalga}) and (\ref{eq:LCalgc}) do vanish.
Therefore the canonical commutators must be modified to:
\BA
&&{\hspace{-0.35in}}{[\, {\tilde J}^+_{\alpha}(x)\,  ,\, {\tilde J}^+_{\beta}(y)\, ]|_{x^+=y^+}}=  i\,\epsilon_{\alpha\beta\gamma} \, {\tilde J}^+_{\gamma}(x)
\delta(x^- - y^-)\,\delta^2(\bit{x}_\perp - \bit{y}_\perp ) \nonumber\\ 
&&\qquad\qquad\qquad\qquad-\oneft i \delta_{\alpha\beta}\partial_-^x\partial_-^y\lbrack S^V\varepsilon(x^- - y^-)\,\delta^2(\bit{x}_\perp - \bit{y}_\perp )\rbrack;
\label{eq:LCalgaMOD}\\
&&{\hspace{-0.35in}}{[\, {\tilde J}^+_{5\alpha}(x)\,  ,\, {\tilde J}^+_{\beta}(y)\, ]|_{x^+=y^+}}= i\,\epsilon_{\alpha\beta\gamma} \, {\tilde J}^+_{5\gamma}(x)
\delta(x^- - y^-)\,\delta^2(\bit{x}_\perp - \bit{y}_\perp)\ ;
\label{eq:LCalgbMOD}\\
&&{\hspace{-0.35in}}{[\, {\tilde J}^+_{5\alpha}(x)\,  ,\, {\tilde J}^+_{5\beta}(y)\, ]|_{x^+=y^+}}= i\,\epsilon_{\alpha\beta\gamma} \, {\tilde J}^+_{\gamma}(x)
\delta(x^- - y^-)\,\delta^2(\bit{x}_\perp - \bit{y}_\perp)\nonumber\\ 
&&\qquad\qquad\qquad\qquad-\oneft i \delta_{\alpha\beta}\partial_-^x\partial_-^y\lbrack S^A\varepsilon(x^- - y^-)\,\delta^2(\bit{x}_\perp - \bit{y}_\perp )\rbrack\ .
\label{eq:LCalgcMOD}
\EA
Here $S^{V,A}$ are taken to be c-numbers, which is the minimal prescription which will satisfy the constraints of Eqs.~(\ref{eq:LCalgaKLa}) and (\ref{eq:LCalgaKLb}).
By matching one finds\footnote{Setting $S_V=S_A$ gives the first spectral function sum rule~\protect\cite{Weinberg:1967kj}. This is expected
since the original derivation obtained this relation by equating the Schwinger terms of the (instant-form) time-space current commutators, which are felt by the $++$
null-plane current commutators at the tip of the light cone.}
\BA
&&{S_V} = \int d\lambda^2 \rho^{(1)}_{V}(\lambda^2)/{\lambda^2}   \label{eq:Schwmatcha} \ ;\\
&&{S_A} = \int d\lambda^2 \left(\rho^{(0)}_{A}(\lambda^2)\,+\, \rho^{(1)}_{A}(\lambda^2)/{\lambda^2} \right)  \ .
\label{eq:Schwmatchb}
\EA
Note that due to parity conservation, the mixed axial-vector commutator has no vacuum expectation value and therefore no non-canonical modification of the current
algebra is necessary.  Clearly the infinite-dimensional symmetry of the canonical current algebra is, in general, modified by these extra terms
that represent vacuum physics at the boundary of the longitudinal coordinate (see fig.~\ref{fig:transplane}).
Integrating over the longitudinal direction, one obtains new purely transverse currents
\begin{eqnarray}
{\tilde F}_{(5)\alpha} (x) \ = \ {\tilde F}_{(5)\alpha} (x^+,\bit{x}_\perp) \ = \ \int\, d x^-\, {\tilde J}^+_{(5)\alpha}(x) 
\label{eq:npchargesBdefined2d}
\end{eqnarray}
which satisfy the current algebra:
\BA
{[\, {\tilde F}_{\alpha}(x)\,  ,\, {\tilde F}_{\beta}(y)\, ]|_{x^+=y^+}}\, &=& \, i\,\epsilon_{\alpha\beta\gamma} \, {\tilde F}_{\gamma}(x)
\delta^2(\bit{x}_\perp - \bit{y}_\perp )\ ;
\label{eq:LCalga2d}\\
{[\, {\tilde F}_{5\alpha}(x)\,  ,\, {\tilde F}_{\beta}(y)\, ]|_{x^+=y^+}}\, &=& \, i\,\epsilon_{\alpha\beta\gamma} \, {\tilde F}_{5\gamma}(x)
\delta^2(\bit{x}_\perp - \bit{y}_\perp)\ ;
\label{eq:LCalgb2d}\\
{[\, {\tilde F}_{5\alpha}(x)\,  ,\, {\tilde F}_{5\beta}(y)\, ]|_{x^+=y^+}}\, &=& \, i\,\epsilon_{\alpha\beta\gamma} \, {\tilde F}_{\gamma}(x)
\delta^2(\bit{x}_\perp - \bit{y}_\perp)\ .
\label{eq:LCalgc2d}
\EA These commutation relations, by construction, have no vacuum
expectation value and satisfy an ${\mathfrak su}(2)_L\otimes {\mathfrak su}(2)_R$
algebra at each point in the transverse plane.  It is this
infinite-dimensional chiral symmetry on the transverse plane that is
exploited in this paper; indeed all of the sum rules that are derived
follow from these brackets. 

One immediately wonders whether, in principle, singular terms
involving purely transverse gradients could appear on the right-hand
side of Eqs.~(\ref{eq:LCalgaMOD}-\ref{eq:LCalgcMOD}).  Such terms
would lead to a breaking of the infinite-dimensional chiral symmetry
on the transverse plane indicated by
Eqs.~(\ref{eq:LCalga2d}-\ref{eq:LCalgc2d}), and leave only the global
chiral symmetry satisfied by the chiral charges that must be present
after integration over the transverse coordinates.  It is
straightforward to construct examples of such singular
terms~\cite{Dicus:1972vp,Pantforder:1997ii}. For instance, consider
the QCD Symanzik action~\cite{Symanzik:1983dc}, a continuum effective field theory of lattice
QCD near the continuum limit, which in general contains a
dimension-five operator, the Pauli term, which scales like the lattice
spacing and gives the quarks a (chromo)magnetic
moment~\cite{Sheikholeslami:1985ij}.  This operator breaks chiral
symmetry in the same way as the quark mass matrix and can be shown to
contribute a transverse gradient to the right hand side of the current
algebra in Eqs.~(\ref{eq:LCalgaMOD}-\ref{eq:LCalgcMOD}).  Of course
this extra term violates scaling and vanishes in the continuum limit.
Here the sum rules that are derived from
Eqs.~(\ref{eq:LCalga2d}-\ref{eq:LCalgc2d}) can be viewed as a means to
experimentally verify the hypothesis that these transverse gradient
terms are not present in QCD.

The decoupling of the physics of the longitudinal dimension is a
remarkable and well-known property of the null-plane
formulation~\cite{Belitsky:2005qn}. It of course leads to the masking
of Lorentz invariance, as discussed in the introduction.  However, the
non-local light-like correlations that are introduced by integrating
out a dimension of space do not lead to conflicts with causality. In
null-plane quantization physics need not be local in $x^--y^-$ as
$(x-y)^2$ does not depend on $x^--y^-$ when $x^+=y^+$ (see footnote 4). Hence,
causality is ensured by locality in the transverse coordinates alone,
as is made manifestly clear in the form of Eqs.~(\ref{eq:LCalgaMOD}-\ref{eq:LCalgcMOD}).

\subsection{Charge and moment algebras}
\label{NPQCDconC2}

\noindent The null-plane chiral symmetry charges are defined by
\begin{eqnarray}
{\tilde Q}_{\alpha} \ = \ \int\, d x^-\, d^2 \bit{x}_\perp\, {\tilde J}^+_{\alpha}(x) \ \ , \ \  {\tilde Q}_{5\alpha}(x^+) \ = \ \int\, d x^-\, d^2 \bit{x}_\perp\, {\tilde J}^+_{5\alpha}(x) \ .
\label{eq:npchargesBdefined}
\end{eqnarray}
The axial-vector charge is null-plane time dependent as the axial-vector current is not conserved. 
These charges generate the ${\mathfrak su}(2)_L\otimes {\mathfrak su}(2)_R$ algebra:
\BA
&& {[\, {\tilde Q}_\alpha\,  ,\, {\tilde Q}_\beta\, ]}\, =\, i\,\epsilon_{\alpha\beta\gamma} \, {\tilde Q}_\gamma \ ; \label{eq:npchargealgA}\\
&& {[\, {\tilde Q}_{5\alpha}(x^+)\, ,\, {\tilde Q}_\beta\, ]}\, =\,  i\,\epsilon_{\alpha\beta\gamma}\,  {\tilde Q}_{5\gamma}(x^+) \ ; \label{eq:npchargealgB}\\ 
&&{[\, {\tilde Q}_{5\alpha}(x^+) \, ,\, {\tilde Q}_{5\beta}(x^+)\, ]} \, = \,  i\,\epsilon_{\alpha\beta\gamma} \, {\tilde Q}_\gamma \ .
\label{eq:npchargealgC}
\EA 
One can also form moments of the currents. Of particular interest are the following:
\begin{eqnarray}
{\tilde {d}}^r_{(5)\alpha} (x^+) \, = \, \int\, d x^-\, d^2 \bit{x}_\perp\, x^r\, {\tilde J}^+_{(5)\alpha}(x) \ \ , \  \
{\tilde {r}}^{rs}_{(5)\alpha} (x^+) \, = \, \int\, d x^-\, d^2 \bit{x}_\perp\, x^r x^s\, {\tilde J}^+_{(5)\alpha}(x) 
\label{eq:npmoments}
\end{eqnarray}
where $r,s=1,2$ are transverse spatial indices. The null-plane time
dependence of the axial charge and of all the higher moments has been
left explicit, as these operators are not conserved~\cite{Beane:2013oia}. This is of crucial importance for
all that follows, as only operators that depend on null-plane time
will have non-vanishing matrix elements between states that exchange
energy. In extracting the physical content from commutators of
operators, matrix elements between physical states must be taken and
the insertion of a complete set of states in the product of operators
will be non trivial only if both operators depend on null-plane
time. Hence, the matrix elements between physical momentum states of
Eq.~(\ref{eq:npchargealgA}) and Eq.~(\ref{eq:npchargealgB}) are
trivial and simply give information about the isospin transformation
properties of matrix elements of charges, whereas the matrix element of
Eq.~(\ref{eq:npchargealgC}) is highly non-trivial and provides
powerful constraints on the manner in which chiral charge spreads out
among the hadrons in the broken phase via pion-hadron scattering.  Other non-trivial commutators
among the moments and charges follow simply from the current algebra,
Eqs.~(\ref{eq:LCalga2d}-\ref{eq:LCalgc2d}). The simplest non-trivial
vector-vector commutator is:
\BA
{[\, {\tilde {d}}^r_\alpha(x^+) \, ,\, {\tilde {d}}^s_\beta(x^+)\, ]} & = &  i\,\epsilon_{\alpha\beta\gamma} \, {\tilde {r}}^{rs}_\gamma(x^+) \ .
\label{eq:npDDR}
\EA 
This commutator provides constraints on the moments of the hadronic vector form factors via Compton scattering.
The simplest non-trivial axial-vector commutator is:
\BA
{[\, {\tilde Q}_{5\alpha}(x^+) \, ,\, {\tilde {d}}^r_\beta(x^+)\, ]} & = &  i\,\epsilon_{\alpha\beta\gamma} \, {\tilde {d}}^r_{5\gamma} (x^+) \ ,
\label{eq:npQDD}
\EA 
which provides constraints on the moments of the hadronic 
axial-vector form factors via pion photoproduction. A second mixed axial-vector commutator,
\BA
{[\, {\tilde Q}_{5\alpha}(x^+) \, ,\, {\tilde {r}}^{rs}_\beta(x^+)\, ]} & = &  i\,\epsilon_{\alpha\beta\gamma} \, {\tilde {r}}^{rs}_{5\gamma} (x^+) \ ,
\label{eq:npQrr}
\EA 
constrains the axial radii of hadrons via pion electroproduction.
This identification of non-trivial Lie brackets continues to hold for
the commutator of any two null-plane-time-dependent operators that are
constructed from the vector and axial-vector charge distributions.  Hence it
is useful to define the general moment operators
\BA
{\cal O}^{\Gamma_{(5)}}_{(5)\alpha}(x^+) \ \equiv \ \int\, d x^-\, d^2 \bit{x}_\perp\, \Gamma_{(5)}\,{\tilde J}^+_{(5)\alpha}(x) \ =\ 
\int\, d^2 \bit{x}_\perp\, \Gamma_{(5)}\,{\tilde F}_{(5)\alpha}(x) \ ,
\label{eq:genmomaxial}
\EA 
where $\Gamma_{(5)}=\Gamma_{(5)}(\bit{x}_\perp)$ is a function of the transverse coordinates and can be a tensor of arbitrary rank. With $\Gamma_5=\Gamma={\bf 1}$, these operators
reduce to the null-plane axial and vector charges. The moment operators then satisfy an algebra that is equivalent to that of Eqs.~(\ref{eq:LCalga2d}-\ref{eq:LCalgc2d})
\BA
{[\, {\cal O}^{\Gamma\;}_{\;\alpha}(x^+)  \,  ,\, {\cal O}^{\bar\Gamma\;}_{\;\beta}(x^+) \, ]}\, &=& \, i\,\epsilon_{\alpha\beta\gamma} {\cal O}^{\Gamma\;{\bar\Gamma}\;}_{\gamma}(x^+) \ ;
\label{eq:momLCalga}\\
{[\, {\cal O}^{\Gamma_5}_{5\alpha}(x^+)  \,  ,\, {\cal O}^{\Gamma\;}_{\;\beta}(x^+) \, ]}\, &=& \, i\,\epsilon_{\alpha\beta\gamma} {\cal O}^{\Gamma_5\Gamma\;}_{5\gamma}(x^+)\ ;
\label{eq:momLCalgb}\\
{[\, {\cal O}^{\Gamma_5}_{5\alpha}(x^+)  \,  ,\, {\cal O}^{{\bar\Gamma}_5}_{5\beta}(x^+) \, ]}\, &=& \, i\,\epsilon_{\alpha\beta\gamma} {\cal O}^{\Gamma_5{{\bar\Gamma}_5}}_{\gamma}(x^+) \ .
\label{eq:momLCalgc}
\EA 
The commutators hold globally and the infinite-dimensionality arises
from the arbitrary choice of the moments via $\Gamma_{(5)}$ and
$\bar\Gamma_{(5)}$. It is straightforward to express the moment
algebra as a closed algebraic system by working in the basis of the
circular harmonics, $e^{i\ell\theta}$, where $\ell$ is the angular momentum on
the plane. An arbitrary tensor in the transverse coordinates,
$x^rx^s\ldots x^t$, can be decomposed into circular harmonics by
contracting with the two-dimensional vectors $\bit{e}^r_\perp = (1,
i)$ and $\bit{\bar e}^r_\perp = (1, -i)$, whose product, $\bit{\bar
  e}^r_\perp \bit{e}^s_\perp =\delta^{rs}+i\epsilon^{rs}$, contains
the two $SO(2)$ tensors.  In circular coordinates, $x^1=r\cos\theta$
and $x^2=r\sin\theta$ where $r = |{\bf x}_\perp|$.  Therefore, the
holomorphic and anti-holomorphic transverse coordinates are $z\equiv
\bit{e}^r_\perp x^r=x^1+ix^2=re^{i\theta}$ and ${\bar z}\equiv \bit{\bar
  e}^r_\perp x^r=x^1-ix^2=re^{-i\theta}$. The moments defined as
\BA
{\cal O}^{m,{\bar m}}_{(5)\alpha}(x^+) \ =\ \int\, d^2 \bit{x}_\perp\, z^m {\bar z}^{\bar m}\,{\tilde F}_{(5)\alpha}(x) \ ,
\label{eq:genalgmom}
\EA 
with $m,{\bar m}\in {\bf Z}$ satisfy the closed algebra
\BA
{[\, {\cal O}^{m,{\bar m}}_{\;\alpha}(x^+)  \,  ,\, {\cal O}^{n,{\bar n}}_{\;\beta}(x^+) \, ]}\, &=& \, i\,\epsilon_{\alpha\beta\gamma} {\cal O}^{m+n,{\bar m}+{\bar n}}_{\gamma}(x^+) \ ;
\label{eq:fullKMa}\\
{[\, {\cal O}^{m,{\bar m}}_{5\alpha}(x^+)  \,  ,\, {\cal O}^{n,{\bar n}}_{\;\beta}(x^+) \, ]}\, &=& \, i\,\epsilon_{\alpha\beta\gamma} {\cal O}^{m+n,{\bar m}+{\bar n}}_{5\gamma}(x^+)\ ;
\label{eq:fullKMb}\\
{[\, {\cal O}^{m,{\bar m}}_{5\alpha}(x^+)  \,  ,\, {\cal O}^{n,{\bar n}}_{5\beta}(x^+) \, ]}\, &=& \, i\,\epsilon_{\alpha\beta\gamma} {\cal O}^{m+n,{\bar m}+{\bar n}}_{\gamma}(x^+) \ .
\label{eq:fullKMc}
\EA
The helicity content of each of the Lie brackets is governed by the
circular harmonic with $\ell=m+n-{\bar m}-{\bar n}$.  The
holomorphic (anti-holomorphic) sector with ${\bar m}={\bar n}=0$ (${m}={n}=0$)
corresponds to the maximal (minimal) helicity. For instance, Eq.~(\ref{eq:npDDR})
is contained in Eq.~(\ref{eq:fullKMa}) with $n=m=1$, ${\bar m}={\bar n}=0$
corresponding to $\ell=2$, $n=m=0$, ${\bar m}={\bar n}=1$
corresponding to $\ell=-2$, and $m={\bar n}=1$, $n={\bar m}=0$
and $m={\bar n}=0$, $n={\bar m}=1$ corresponding to $\ell=0$.

Evidently the algebra of Eqs.~(\ref{eq:fullKMa}-\ref{eq:fullKMc})
corresponds to a mapping of the transverse plane to the chiral group
${SU}(2)_L\otimes {SU}(2)_R$. It is important to stress that, despite
the presence of the algebra, only the isospin charges are conserved
quantities. It is the null-plane time dependence of the operators
which give rise to non-trivial constraints on S-matrix elements.  In
the limit of unit radius, ${\bar z} = z^{-1}$, this algebra reduces to
the mapping of the unit circle to the chiral group and is thus an
untwisted affine Kac-Moody algebra without central charge (loop
algebra), ${\widehat{\mathfrak{s} \mathfrak{u}}}(2)_L\otimes
{\widehat{\mathfrak{s}
    \mathfrak{u}}}(2)_R$~\cite{Goddard:1986bp,kac;1994}. The moments
with unit radius are unrelated to forward scattering.

Extracting information from the moments of the currents rather than
the currents themselves is a matter of choice. Indeed all of the sum
rules that follow from the moment algebra can be obtained by
considering matrix elements of the commutators of currents (null-plane
time and space components) and then Taylor expanding in the momentum
transfer variable.  We choose to work with the moments as they are
well-defined objects that obey a hierarchy of scales on the transverse
plane and satisfy a non-singular closed algebra. In addition, there is
a subtle commutation of limits in working directly with the currents
which we choose to avoid~\cite{Dicus:1971uk}.

\subsection{Fundamental matrix element}
\label{fme}

\begin{figure}[!t]
  \centering
     \includegraphics[scale=0.24]{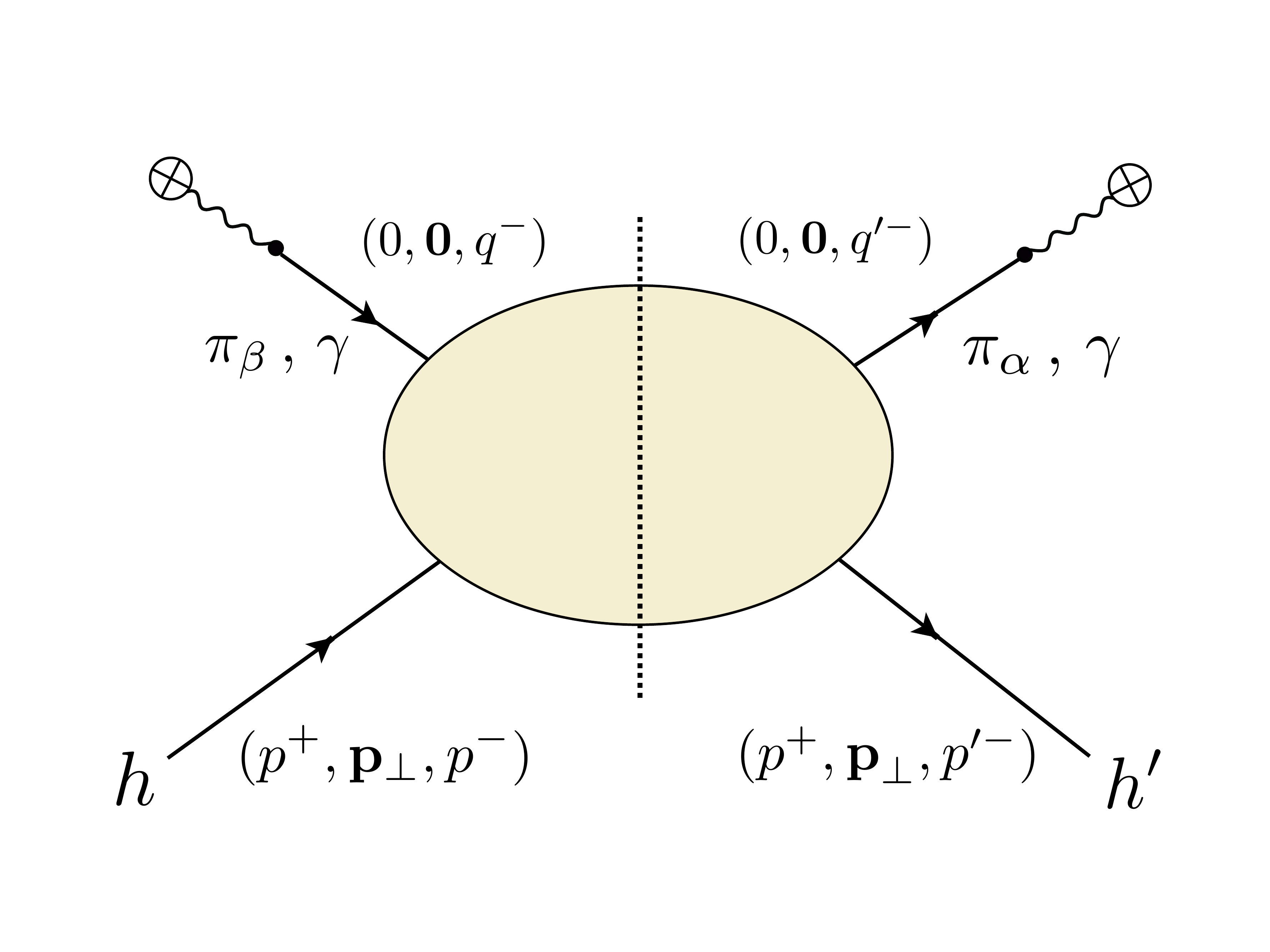}
     \caption{Basic forward kinematics of the moment sum rules. The vertical dashed line indicates that only physical states are present as intermediate states. The process therefore describes 
a pion ($\pi^\alpha$) or an isovector photon ($\gamma$) scattering in the forward direction from a hadronic target $h$ to give a pion or an isovector photon and a hadron $h^\prime$.}
  \label{fig:scatt}
\end{figure}

\noindent  In order to extract physics from the Lie brackets satisfied by the moments, one
must take matrix elements of the commutators between hadron states $h'$ and $h$~\footnote{It is understood that these matrix elements will be non-vanishing only if 
allowed by QCD symmetries. For instance, if $h$ is a baryon, then so is $h'$.}. 
The fundamental matrix element to be evaluated
is:
\BA
&&\langle\, h'\,,\,\lambda'\,;\,p^{\prime +}\,,{\bf p}_\perp'\, |\, {\cal O}^{\Gamma_{(5)}}_{(5)\alpha}(x^+) \, {\cal O}^{{\bar\Gamma}_{(5)}}_{(5)\beta}(x^+)\, |\, h \,,\,\lambda\,;\,p^+\,,\,{\bf p}_\perp\,\rangle   \ .
\label{eq:fundmatelem} 
\EA 
This matrix element of the product of moment operators is exhibited
pictorially in fig.~\ref{fig:scatt}. Details regarding the null-plane
momentum states are relegated to Appendix~\ref{NPQCDconE}. While
general sum rules will be derived for arbitrary structure functions,
the choice of pions coupling to the axial-vector current and isovector
photons coupling to the vector current will then be made, both
impinging on nucleon targets. As will be seen in detail, the blob in fig.~\ref{fig:scatt}
corresponds to the imaginary part of the forward scattering amplitude
(and its derivatives) for pion scattering, Compton scattering and pion
photo/electroproduction.
In the forward limit ($t=0$), we choose a coordinate system with
four-momenta assigned as in fig.~\ref{fig:scatt}: $q=(0,{\bf 0},q^-)$,
$q'=(0,{\bf 0},q^{\prime -})$, $p=(p^+,{\bf p}_\perp,p^{-})$, and
$p'=(p^+,{\bf p}_\perp,p^{\prime -})$. Therefore, the Mandelstam
variables are:
\BA
s &=& (p\,+\,q)^2 \ =\ M_h^2 \ +\ 2p^+ q^- \\
  &=& (p'\,+\,q')^2 \ =\ M_{h'}^2 \ +\ 2p^+ q^{\prime -} \ ; \\
u &=& (p'\,-\,q)^2 \ =\ M_{h'}^2 \ -\ 2p^+ q^- \\
  &=& (p\,-\,q')^2 \ =\ M_{h}^2 \ -\ 2p^+ q^{\prime -} \ ,
\label{eq:mandelstam}
\EA
and ${\bar\nu}\equiv p\cdot q=p^+q^-$. Off-forward kinematics will be
adopted in the derivation of the sum rule in Section~\ref{av2}, using the
kinematical choices outlined in Appendix~\ref{sec:ScattConv} (Photo/electroproduction).

Generally, a current carrying four-momentum $q$ carries a squared mass
given by $q^2=2q^+q^--{\bf q}_\perp^2$. Hence, by choosing $q^+=0$,
one can hold the mass fixed at its space-like value while integrating over the null-plane
energy $q^-$\footnote{This ``fixed-mass'' limit has nothing to do with short-distance physics and is
therefore far from the kinematical region where QCD degrees of freedom may be profitably used. Indeed, this
limit is naturally related to a long-distance multipole-like expansion on the transverse plane, as will be
seen below.}. 
This kinematical property of the null-plane formulation is at the heart of its utility. For instance,
it is precisely this property which leads to Goldstone's Theorem on the null-plane~\cite{Beane:2013oia}.
As our currents are massless, our treatment of $q$ and $q'$ as ``null-vectors'' with vanishing momentum 
facilitates the derivation of the sum rules. 

The matrix element Eq.~(\ref{eq:fundmatelem}) can be evaluated by inserting a complete set of states between the moment operators.
This sum contains the infinite number of single-particle and continuum states that are connected to the target hadrons
by the moment operator.  It is convenient to remove the ``Born'' contributions, where the moment
operators are connecting hadron states to themselves. The relevant matrix elements are:
\BA
&&{\hspace{-0.41in}}\langle\, h'\,,\,\lambda'\,;\,p'\, | {\cal O}^{\Gamma_{(5)}}_{(5)\alpha}(x^+) |\, h \,,\,\lambda\,;\,p\,\rangle   
\, = \,  \int\, d x^-\, d^2 \bit{x}_\perp\, \Gamma_{(5)}\, e^{ix\cdot q}
\langle\, h'\,,\,\lambda'\,;\,p'\, | {\tilde J}_{(5)\alpha}^+(0) |\, h \,,\,\lambda\,;\,p\,\rangle  ,
\label{eq:momBorn}
\EA 
where translational invariance (see  Eq.~(\ref{eq:OLI})) has been used and here $q=(p^{\prime+}-p^+, {\bf p}_\perp'-{\bf p}_\perp,q^{-})$. Therefore the Born contributions ($h=h'$) will 
be given by the elastic vector and axial-vector form factors of the hadron $h$ and their derivatives in the forward direction.

Contributions from all other states that appear in the completeness sum are efficiently extracted by considering
the null-plane time dependence of the moment operators. Using Eq.~(\ref{eq:gentdep}) gives
\BA
&&{\hspace{-0.45in}}{\cal O}^\Gamma_{\alpha}(x^+)\, |h\, ,\,\lambda\,;\, p\,\rangle \, = \, \frac{i}{\left(p^- - P^-\right)}
\int d x^- d^2 \bit{x}_\perp\, \left(\partial^r\Gamma\right)\,{\tilde J}^r_{\alpha}(x) |h\, ,\,\lambda\,;\, p\,\rangle \, ;
\label{eq:genmomvectorME} \\
&&{\hspace{-0.45in}}{\cal O}^{\Gamma_5}_{5\alpha}(x^+)\, |h\, ,\,\lambda\,;\, p\,\rangle  \, =\,  \frac{i}{\left(p^- - P^-\right)}
\int d x^- d^2 \bit{x}_\perp\, \big\lbrack\, \left(\partial^r\Gamma_5\right)\,{\tilde J}^r_{5\alpha}(x) 
 + \Gamma_5  {\tilde {\cal D}}_{5\alpha}(x) \, \big\rbrack |h\, ,\,\lambda\,;\, p\,\rangle  .
\label{eq:genmomaxialME}
\EA 
Note the appearance of the energy denominators which include the null-plane Hamiltonian operator $P^-$. 
These are critical in what follows as they give the appropriate energy weightings of the sum rules which
are necessary for convergence in the high-energy limit.

There is only a single conserved moment, ${\cal
  O}^{1}_{\alpha}={\tilde {Q}}_\alpha$, which is the isovector
charge, whose non-vanishing matrix elements give the isospin matrix of
the hadron,
\BA
\langle\, h'\,,\,\lambda'\,;\,p'\,|{\tilde {Q}}_\alpha|\, h \,,\,\lambda\,;\,p\,\rangle & =& 
 \,(2\pi)^3\,\delta(\,q^+\,)\,\delta^2(\,{\bf q}_\perp \,)\langle\, h'\,,\,\lambda'\,;\,p'\,|{\tilde J}^+_\alpha (0)|\, h \,,\,\lambda\,;\,p\,\rangle \nonumber \\
&=&  \,(2\pi)^3\,2\,p^+\,\delta(\,q^+\,)\,\delta^2(\,{\bf q}_\perp \,)\lbrack\, T_\alpha \,\rbrack_{h}\, \delta_{h' h}\,\delta_{\lambda'\lambda} \ .
\label{eq:isodef}
\EA 
By contrast, the matrix element of the axial charge is an S-matrix element,
\BA
\langle\, h'\, ,\,\lambda'\, |\, {\tilde Q}^5_\alpha(x^+) \,|\, h\, ,\,\lambda\, \rangle \, =\, 
(2\pi)^3\,2\,p^+\,\delta(\,q^+\,)\,\delta^2(\,{\bf q}_\perp \,)
\lbrack\, X_\alpha(\lambda) \,\rbrack_{h' h}\,
\delta_{\lambda'\lambda} \ ,
\label{eq:HME40}
\EA
where $\lbrack\, X_\alpha(\lambda) \,\rbrack_{h' h}$ is the hadronic axial-vector coupling matrix. 
In particular, the Feynman amplitude for the pion transition process $h\rightarrow h' +\pi$  may be written~\cite{Weinberg:1969hw}
\BA
{\cal M}_\alpha (\,p',\,\lambda',\, h'\,;\,p,\,\lambda,\, h\,) \ =\ \frac{i}{F_\pi}\,
(\,M^2_h\ -\ M^2_{h'}\,)\, \lbrack\, X_\alpha(\lambda) \,\rbrack_{h' h}\,
\delta_{\lambda'\lambda} \ . 
\label{eq:HME42}
\EA

%%%%%%%%%%%%%%%%%%%%%%%%%%%%%%%%%%%%%%%%%%%%%%%%%%%%%%%%%%%%%%%%%%%%%%%%%%%%%%%%%%%%%%%%%%%%%%%%%%%%%%%%%%%%%%%%%%%%%%%%%%%%%%%%%%%%%%%%
\section{Moment sum rules}
\label{msr}

%%%%%%%%%%%%%%%%%%%%%%%%%%%%%%%%%%%%%%%%%%%%%%%%%%%%%%%%%%%%%%%%%%%%%%%%%%%%%%%%%%%%%%%%%%%%%%%%%%%%%%%%%%%%%%%%%%%%%%%%%%%%%%%%%%%%%%%%
\subsection{Axial-Axial}
\label{aa}

%%%%%%%%%%%%%%%%%%%%%%%%%%%%%%%%%%%%%%%%%%%%%%%%%%%%%%%%%%%%%%%%%%%%%%%%%%%%%%%%%%%%%%%%%%%%%%%%%%%%%%%%%%%%%%%%%%%%%%%%%%%%%%%%%%%%%%%%
\subsubsection*{General case}
\label{sec:aagc}

\noindent We will begin with the simplest case.
Consider the matrix element between hadronic momentum states of the commutator of axial charges, Eq.~(\ref{eq:npchargealgC}):
\BA
\langle\, h'\,,\,\lambda\,;\,p'\, | {[\, {\tilde {Q}}_{5\alpha}(x^+) \, ,\, {\tilde {Q}}_{5\beta}(x^+)\, ]}|\, h \,,\,\lambda\,;\,p\,\rangle   
& = &  i\,\epsilon_{\alpha\beta\gamma} \, \langle\, h'\,,\,\lambda\,;\,p'\,|{\tilde {Q}}_\gamma|\, h \,,\,\lambda\,;\,p\,\rangle \ .
\label{eq:MEq5q5}
\EA 
Note that because the charges are Lorentz scalars, they do not change the helicity of the hadronic states (they contain only the $\ell=0$ circular
harmonic). Therefore, $\Delta\lambda \equiv \lambda'-\lambda=\ell=0$.
The Born contribution to the left hand side (LHS) of Eq.~(\ref{eq:MEq5q5}) is:
\BA
&&{\hspace{-0.34in}}\frac{1}{2p^+}
(2\pi)^3\delta\left({q}^+\right)\delta^2\left({\bf q}_\perp\right)\sum_{h''}\langle h'\,,\,\lambda\,;\,p'| {\tilde J}_{5\alpha}^+(0) | h''\,,\,\lambda\,;\,p\rangle 
\langle h''\,,\,\lambda\,;\,p| {\tilde J}_{5\beta}^+(0) | h\,,\,\lambda\,;\,p\rangle 
-{\it c.t.}
\label{eq:bornpih} 
\EA
where ${\it c.t.}$ refers to the second term in the commutator.
The final results are independent of null-plane time and therefore $x^+=0$ is taken from the outset.
Using Eq.~(\ref{eq:genmomaxialME}) and translational invariance, gives for the continuum
part of the LHS of Eq.~(\ref{eq:MEq5q5}):
\begin{eqnarray}
{\hspace{-0.17in}}(2\pi)^3\delta\left({q}^+\right)\delta^2\left({\bf q}_\perp\right) \langle h'\,,\,\lambda |(2\pi)^3
\frac{{\tilde {\cal D}}_{5\alpha}(0)\delta\left(P^+-{p^+}\right)\delta^2\left({\bf P}_\perp-{{\bf p}_\perp}\right)    {\tilde {\cal D}}_{5\beta}(0)}
{\left(P^-\,-\,p^{\prime-}\right)\left(P^-\,-\,p^-\right)}| h \,,\,\lambda\rangle  
 - {\it c.t.}\ .
\label{eq:q5q5contdef} 
\end{eqnarray}
Defining the structure function
\BA 
w_{\lambda;\alpha\beta}^{h' h}(p,q)  & \equiv&
\langle\, h'\,,\,\lambda\,;\,p\, |\,(2\pi)^3 \, {\tilde {\cal D}}_{5\alpha}(0) 
\delta^4\left({q}\,+\,{p}\,-{P}\right){\tilde {\cal D}}_{5\beta}(0)\, |\, h\,,\,\lambda\,;\,p\, \rangle \ ,
\label{eq:wpipidefined} 
\EA
where here $q=(0,{\bf 0},q^-)$, Eq.~(\ref{eq:q5q5contdef}) can be expressed as
\begin{eqnarray}
{\hspace{-0.13in}}(2\pi)^32p^+\delta\left({q}^+\right)\delta^2\left({\bf q}_\perp\right) 
2\int_{{\bar\nu}_T}^\infty\frac{d{\bar\nu}}{{\bar\nu}\left(2{\bar\nu}+M_h^2-M_{h'}^2\right)}{w}_{\lambda; [\alpha\beta]}^{h'h}(p,q) 
\end{eqnarray}
where ${\bar\nu}_T$ is the threshold energy and
\BA
w^{h'h}_{\lambda;[\alpha\beta]}(p,q) \ \equiv\ \oneht \left( w^{h'h}_{\lambda;\alpha\beta}(p,q) \ -\ w^{h'h}_{\lambda;\beta\alpha}(p,q) \right)\ .
\label{eq:wpipiodddefined}
\EA
The RHS of Eq.~(\ref{eq:MEq5q5}) is
\BA
 \,(2\pi)^3\,\delta(\,q^+\,)\,\delta^2(\,{\bf q}_\perp \,)
i\,\epsilon_{\alpha\beta\gamma}\,\langle\, h'\,,\,\lambda\,;\,p'\,|{\tilde J}^+_\gamma (0)|\, h \,,\,\lambda\,;\,p\,\rangle
\label{eq:MEisocharge}
\EA
where Eq.~(\ref{eq:isodef}) has been used. 
Finally, matching the LHS and the RHS, and integrating over the target hadron momenta to remove the momentum delta functions, gives the general form of the structure-function sum rule:
\BA
&&{\hspace{-0.25in}} 2\int_{{{\bar\nu}_T}}^\infty\frac{d{\bar\nu}}{{\bar\nu}\left(2{\bar\nu}+M_h^2-M_{h'}^2\right)} \;
{w}_{\lambda;[\alpha\beta]}^{h'h}(p,q)|_{q^+={\bf q}_\perp =0}
\nonumber \\
&& \ +\ \frac{1}{(2p^+)^2}\sum_{h''}\Big\lbrack
\langle\, h'\,,\,\lambda\,;\,p'\,|\, {\tilde J}_{5\alpha}^+(0)  \, |\, h'' \,,\,\lambda\,;\,p\,\rangle 
\langle\, h''\,,\,\lambda\,;\,p\,|\, {\tilde J}_{5\beta}^+(0)  \, |\, h \,,\,\lambda\,;\,p\,\rangle 
- {\it c.t.} \Big\rbrack|_{q^+={\bf q}_\perp =0} \nonumber \\
&&\quad \ =\ \frac{1}{2p^+}i\,\epsilon_{\alpha\beta\gamma}\,
\,\langle\, h'\,,\,\lambda\,;\,p'\,|{\tilde J}^+_\gamma (0)|\, h \,,\,\lambda\,;\,p\,\rangle
\label{eq:genAWSR}
\EA
Whether the Born term exists or is removed from the integral depends on
the particular hadronic targets. The Lorentz-invariant manner of the derivation and integration
variables ensure that the sum rule is frame-independent.  It is clear
that the structure function is related to the imaginary part of the
$\pi h\rightarrow \pi h'$ forward scattering amplitude. In the case
$h=h'$, the imaginary part of the forward scattering amplitude may be
replaced with the corresponding total $\pi h$ scattering
cross-section via the optical theorem, as will be seen below in the nucleon case.

%%%%%%%%%%%%%%%%%%%%%%%%%%%%%%%%%%%%%%%%%%%%%%%%%%%%%%%%%%%%%%%%%%%%%%%%%%%%%%%%%%%%%%%%%%%%%%%%%%%%%%%%%%%%%%%%%%%%%%%%%%%%%%%%%%%%%%%%
\subsubsection*{Nucleon sum rule}
\label{aansr}

\noindent Here we specialize to the nucleon, $h',h=N$. Using the null-plane decomposition of the nucleon
form factors given in Appendix~\ref{sec:NFFC}, we have 
\BA
\langle\, N\,,\,\lambda\,;\,p'\,|{\tilde J}^+_\alpha (0)|\, N \,,\,\lambda\,;\,p\,\rangle \ =\ 2p^+ \xi^T_a (T_\alpha)_{ab}\xi_b  \ =\ 2p^+ \xi^T\,{F}^V_{1\alpha}(0)\,\xi \ \equiv \ 2p^+ \lbrack\, T_\alpha \,\rbrack_{N} \ ,
\label{eq:MEisocharge2}
\EA
where $\xi_a$ ($a=1,2$) is the nucleon isospinor, and the Born contribution is 
\BA
(2\pi)^3{2p^+}\delta\left({q}^+\right)\delta^2\left({\bf q}_\perp\right)
\xi^T\,[\,{G}^A_{\alpha}(0)\,,\, {G}^A_{\beta}(0)\,]\,\xi \ .
\label{eq:bornpiN} 
\EA
With the scattering amplitude conventions of Appendix~\ref{sec:ScattConv}, 
the sum rule becomes
\BA
\frac{F_\pi^2}{\pi}\int_0^\infty\frac{d{\bar\nu}}{{{\bar\nu}^2}}\,  {\rm Im}\; {\cal T}_{[\alpha\beta]}({\bar\nu},0)  \ +\ 
\xi^T\,[\,{G}^A_{\alpha}(0)\,,\, {G}^A_{\beta}(0)\,]\,\xi 
\ =\ i\,\epsilon_{\alpha\beta\gamma}\,\xi^T\,{F}^V_{1\gamma}(0)\,\xi \ .
\label{eq:genAWSRmu}
\EA
Note that the dependence on helicity has been removed as the sum rule holds for $\lambda=\pm 1/2$.
In this form, the sum rule is particularly physically intuitive. It shows that in the presence
of spontaneous chiral symmetry breaking, the axial-vector charge is not conserved and therefore the axial form factor at zero momentum transfer deviates
from its unbroken value by the axial-vector charge induced matrix elements that connect the nucleon to all other hadronic
states. Using the conventions of Appendix~\ref{sec:ScattConv}, the sum rule can be expressed for $\pi p$ scattering in the variable $\nu$ as
\BA
\frac{4F_\pi^2}{\pi}\int_0^\infty\frac{d\nu}{{\nu^2}}\,{\rm Im}\; D^{-}(\nu,0) \ +\ g_A^2 \ =\ 1 \ ,
\label{eq:NAWSR}
\EA
and finally, using the optical theorem,
\BA 
\boxed{ {{2F_\pi^2}\over\pi}\int_{{\bar\nu}_T}^\infty {{d{\bar\nu}}\over{\bar\nu}}\;
\Big\lbrack\sigma^{\pi^- p}({\bar\nu} )-\sigma^{\pi^+ p}({\bar\nu} )\Big\rbrack \ +\  g_A^2 \ =\ 1 \ .}
\label{eq:AWsumrule}
\EA This is the AW sum rule for pion-nucleon scattering in the chiral
limit~\cite{Adler:1965ka,Weisberger:1965hp}.  Here ${\bar\nu_T}$ is
the physical threshold. Without the benefit of $\chi$PT, the original
papers made various attempts at extrapolating away from the chiral
limit (see also Ref.~\cite{Brown:1971pn}).  The leading, universal
chiral corrections to the sum rule are obtained in
Ref.~\cite{Klco:2015} using $\chi$PT, which in addition performs an
updated analysis of the sum rule and finds excellent agreement with
experiment.

%%%%%%%%%%%%%%%%%%%%%%%%%%%%%%%%%%%%%%%%%%%%%%%%%%%%%%%%%%%%%%%%%%%%%%%%%%%%%%%%%%%%%%%%%%%%%%%%%%%%%%%%%%%%%%%%%%%%%%%%%%%%%%%%%%%%%%%%
\subsection{Vector-Vector}
\label{vv}

%%%%%%%%%%%%%%%%%%%%%%%%%%%%%%%%%%%%%%%%%%%%%%%%%%%%%%%%%%%%%%%%%%%%%%%%%%%%%%%%%%%%%%%%%%%%%%%%%%%%%%%%%%%%%%%%%%%%%%%%%%%%%%%%%%%%%%%%
\subsubsection*{General case}
\label{vvgc}

\noindent We will proceed with a derivation of the sum rules relevant to forward Compton scattering.
This derivation parallels that of the last section however it is complicated by the presence of the higher moments.
Consider the matrix element between hadronic momentum states of the commutator, Eq.~(\ref{eq:npDDR}):
\BA
\langle\, h'\,,\,\lambda'\,;\,p'\, | {[\, {\tilde {d}}^r_\alpha(x^+) \, ,\, {\tilde {d}}^s_\beta(x^+)\, ]}|\, h \,,\,\lambda\,;\,p\,\rangle  & = &  i\,\epsilon_{\alpha\beta\gamma} \, \langle\, h'\,,\,\lambda'\,;\,p'\, |{\tilde {r}}^{rs}_\gamma(x^+)|\, h \,,\,\lambda\,;\,p\,\rangle \ .
\label{eq:MEnpDDR}
\EA 
As has been shown above, this commutator can be decomposed into circular harmonics $\ell=0,\pm 2$.
Therefore, in general there are non-trivial sum rules for $\Delta\lambda =\ell=0,\pm 2$.
As in the previous case, the Born contribution will first be extracted from the LHS. One finds
\BA
{\hspace{-0.25in}}\langle\, h'\,,\,\lambda'\,;\,p'\,|\,  {\tilde {d}}^{r}_\alpha(x^+) \, |\, h \,,\,\lambda\,;\,p\,\rangle  =
(2\pi)^3 \delta(\,{q^+}\,)\lbrack\, i\partial^r_{{\bf p}^\prime}\,\delta^2(\,{\bf q_\perp}\,)\,\rbrack 
\langle\, h'\,,\,\lambda'\,;\,p'\,|\, {\tilde J}_\alpha^+(0)  \, |\, h \,,\,\lambda\,;\,p\,\rangle \; ,
\label{eq:matd} 
\EA
where we are using the notation $\partial^r_{\bf p}\equiv\partial/\partial p^r$.
And the Born contribution to the LHS is
\BA
&&{\hspace{-0.5in}}\frac{1}{2p^+} (2\pi)^3 \delta(\,{q^+}\,)
\int\, d^2 {\bf p}^{\prime\prime}_\perp\,
\lbrack\, i\partial^r_{{\bf p}^\prime}\,\delta^2(\,{\bf p}^{\prime}_\perp-{\bf p}^{\prime\prime}_\perp \,)\,\rbrack 
\lbrack\, i\partial^s_{{\bf p}^{\prime\prime}}\,\delta^2(\,{\bf p}^{\prime\prime}_\perp-{\bf p}_\perp \,)\,\rbrack \nonumber \\
&&{\hspace{-0.5in}}\times\, \sum_{h''}
\langle\, h'\,,\,\lambda'\,;\,p'\,|\, {\tilde J}_\alpha^+(0)  \, |\, h'' \,,\,\lambda''\,;\,p''\,\rangle 
\langle\, h''\,,\,\lambda''\,;\,p''\,|\, {\tilde J}_\beta^+(0)  \, |\, h \,,\,\lambda\,;\,p\,\rangle 
- {\it c.t.} \ .
\label{eq:born1} 
\EA
The RHS of Eq.~(\ref{eq:MEnpDDR}) is
\BA
(2\pi)^3 \delta(\,{q^+}\,)\lbrack\, i\partial^r_{\bf p}i\partial^s_{\bf p}\,\delta^2(\,{\bf q}_\perp\,)\,\rbrack 
i\,\epsilon_{\alpha\beta\gamma} \langle\, h'\,,\,\lambda'\,;\,p'\,|\, {\tilde J}_\alpha^+(0)  \, |\, h \,,\,\lambda\,;\,p\,\rangle \ .
\label{eq:vvrhs} 
\EA
Proceeding as in the axial-axial case, the continuum contribution to the LHS is
\BA{\hspace{-0.5in}}
(2\pi)^3 \delta(\,q^+\,)\,\delta^2(\,{\bf q}_\perp \,)
\langle h'\,,\,\lambda' |(2\pi)^3
\frac{{\tilde J}_{\alpha}^r(0)\delta\left(P^+-{p^+}\right)\delta^2\left({\bf P}_\perp-{{\bf p}_\perp}\right)    {\tilde J}_{\beta}^s(0)}
{\left(P^-\,-\,p^{\prime-}\right)\left(P^-\,-\,p^-\right)}| h \,,\,\lambda\rangle  
- {\it c.t.} \ .
\label{eq:MEnpRHS3}
\EA
Now defining the structure function
\BA {\hspace{-0.3in}}
w_{\lambda'\lambda;\alpha\beta}^{h'h;rs}(p,q)& \equiv&
\langle\, h'\,,\,\lambda'\,;\,p\, |\,(2\pi)^3 \, {\tilde J}_{\alpha}^r(0) 
\delta^4\left({q}\,+\,{p}\,-{P}\right){\tilde J}_{\beta}^s(0)\, |\, h\,,\,\lambda\,;\,p\, \rangle \ ,
\label{eq:wgamgamdefined} 
\EA
and after undergoing various standard manipulations, the general form of the structure-function sum rule is found to be
\BA
&&\int_{{\bar\nu}_T}^\infty\frac{d{\bar\nu}}{{\bar\nu}\left(2{\bar\nu}+M_h^2-M_{h'}^2\right)}\left(w_{\lambda'\lambda;\alpha\beta}^{h'h;rs}(p,q) -w_{\lambda'\lambda;\beta\alpha}^{h'h;sr}(p,q)\right)|_{q^+={\bf q}_\perp=0}  \nonumber \\
&&+\frac{1}{(2p^+)^2}\Big\lbrace\,\sum_{h''}  
i\partial^r_{{\bf p}^\prime}i\partial^s_{{\bf p}^\prime}\,\lbrack 
\langle\, h'\,,\,\lambda'\,;\,p'\,|\, {\tilde J}_\alpha^+(0)  \, |\, h'' \,,\,\lambda''\,;\,p\,\rangle 
\langle\, h''\,,\,\lambda''\,;\,p\,|\, {\tilde J}_\beta^+(0)  \, |\, h \,,\,\lambda\,;\,p\,\rangle  \,\rbrack |_{q^+={\bf q}_\perp=0}\nonumber \\
&&+\sum_{h''}\Big\lbrack  
i\partial^r_{{\bf p}^\prime}\left( i\partial^s_{\bf k}\lbrack 
\langle\, h'\,,\,\lambda'\,;\,p'\,|\, {\tilde J}_\alpha^+(0)  \, |\, h'' \,,\,\lambda''\,;\,k\,\rangle 
\langle\, h''\,,\,\lambda''\,;\,k\,|\, {\tilde J}_\beta^+(0)  \, |\, h \,,\,\lambda\,;\,p\,\rangle  \,\rbrack\right)|_{k=p}\Big\rbrack |_{q^+={\bf q}_\perp=0}\nonumber \\
&&
- {\it c.t.} \ \Big\rbrace
\ =\  
\frac{1}{2p^+}i\epsilon_{\alpha\beta\gamma} \left(i\partial^r_{{\bf p}^\prime}i\partial^s_{{\bf p}^\prime}\langle\, h'\,,\,\lambda'\,;\,p'\,|\, {\tilde J}_\gamma^+(0)  \, |\, h \,,\,\lambda\,;\,p\,\rangle \right) |_{q^+={\bf q}_\perp=0} \ .
\label{eq:VVsumrulegeneral}
\EA
With isovector photons coupled to the vector currents, it is clear that the structure function can be related to the imaginary part of the forward Compton scattering amplitude.

%%%%%%%%%%%%%%%%%%%%%%%%%%%%%%%%%%%%%%%%%%%%%%%%%%%%%%%%%%%%%%%%%%%%%%%%%%%%%%%%%%%%%%%%%%%%%%%%%%%%%%%%%%%%%%%%%%%%%%%%%%%%%%%%%%%%%%%%
\subsubsection*{Nucleon sum rules}
\label{vvnsr}

\noindent As the nucleon is a spin one-half object, there is no spin
flip sum rule. Therefore as in the previous case, the helicity label
will be omitted.  Making use of the Compton scattering conventions of
Appendix~\ref{sec:ScattConv}, and noting that the first Born term
cancels against the RHS, one readily finds
\BA
{\hspace{-0.5in}}&&\frac{1}{2}\int_{-\infty}^{\infty}\frac{d{\bar\nu}}{{\bar\nu}^2} {\cal C}_{\alpha\beta}^{rs}(p,q) |_{q^+={\bf q}_\perp=0} \ =\  \xi^T\,\Bigg\lbrack 
\left(\,2[\,{F}^{V\prime}_{1\alpha}(0)\,,\, {F}^V_{1\beta}(0)\,]\ -\ \Big\lbrack\,\frac{{F}^V_{2\alpha}(0)}{2m_N}\,,\, \frac{{F}^V_{2\beta}(0)}{2m_N}\,\Big\rbrack\,\right)\,\delta^{rs} \nonumber \\
&&\qquad\qquad\qquad\qquad\qquad\qquad\qquad\qquad\qquad -\ i \Big\lbrace\,\frac{{F}^V_{2\alpha}(0)}{2m_N}\,,\, \frac{{F}^V_{2\beta}(0)}{2m_N}\,\Big\rbrace\,\epsilon^{rs} \Bigg\rbrack \xi \ .
\label{eq:VVsumrulenucleon}
\EA
Note that crossing symmetry has been used to extend the range of integration.
Now the parts symmetric and antisymmetric in the isospin indices can be matched using the
Compton amplitude decomposition given in Appendix~\ref{sec:ScattConv}. 

For the antisymmetric part one finds,
\BA
{\hspace{-0.5in}}&&\int_{{\bar\nu}_T}^\infty {d{\bar\nu}}\;{W}_{1[\alpha\beta]}({\bar\nu},0) \ =\ -\xi^T\,
\left(\,2[\,{F}^{V\prime}_{1\alpha}(0)\,,\, {F}^V_{1\beta}(0)\,]\ -\ \Big\lbrack\,\frac{{F}^V_{2\alpha}(0)}{2m_N}\,,\, \frac{{F}^V_{2\beta}(0)}{2m_N}\,\Big\rbrack\;\right)
\, \xi \ .
\label{eq:CRnucleonampA}
\EA
Using the optical theorem (see Appendix~\ref{sec:ScattConv}) this sum rule can be expressed in terms of a difference of total cross-sections for isovector photons:
\BA
\boxed{{ \frac{1}{3}} \vev{r^2}^V_1 \ =\ \left( \frac{\kappa_V}{2m_N}\right)^2\ +\ \frac{1}{2\pi^2\alpha}\int_{{\bar\nu}_T}^\infty \frac{d{\bar\nu}}{\bar\nu}
\left(\, 2\sigma^V_{1/2}({\bar\nu})\ -\ \sigma^V_{3/2}({\bar\nu})\, \right)\ .} 
\label{eq:CRSR} 
\EA 
This is the Cabbibo-Radicatti (CR) sum
rule~\cite{Cabibbo:1966zz}. Here $\sigma^V_I$ is the total cross-section for an isovector photon
scattering from a proton to a final hadronic state with isospin $I$.
Early analyses of this sum rule were given in
Ref.~\cite{Gilman:1966zz,Adler:1966gd,Dominguez:1975zw} which
confirmed the sum rule at the ten-percent level. Given the significant
recent experimental progress in measuring the various photoproduction
multipoles that dominate this sum rule, it would be interesting to
perform an updated analysis of the CR sum rule.

For the symmetric part one finds,
\BA
&&\int_{{\bar\nu_T}}^\infty \frac{d{\bar\nu}}{\bar\nu}\big\lbrack  {W}_{3(\alpha\beta)}({\bar\nu},0)\ +\ {\bar\nu} {W}_{4(\alpha\beta)}({\bar\nu},0) \big\rbrack\ =\ {\oneht} \xi^T\,
\Big\lbrace\,\frac{{F}^V_{2\alpha}(0)}{2m_N}\,,\, \frac{{F}^V_{2\beta}(0)}{2m_N}\,\Big\rbrace\,  \xi \ .
\label{eq:CRnucleonampS}
\EA
Again using results from Appendix~\ref{sec:ScattConv} the sum rule can be expressed in terms of a difference of total cross-sections:
\BA
\boxed{ \left( \frac{\kappa_V}{2m_N}\right)^2\ =\ \frac{1}{2\pi^2\alpha}\int_{{\bar\nu}_T}^\infty \frac{d{\bar\nu}}{\bar\nu}\left(\, \sigma^V_{P}({\bar\nu})\ -\ \sigma^V_{A}({\bar\nu})\, \right)\ .} 
\label{eq:GDHSR} 
\EA 
This is the (isovector) GDH sum
rule~\cite{Gerasimov:1965et,Drell:1966jv}~\footnote{Note that the
  isoscalar sum rule as well as the GDH sum rules for the nucleons
  themselves are obtained by choosing the appropriate
  currents.}. 
Here $\sigma^V_P$
($\sigma^V_A$) is the total cross-section for an isovector photon
scattering from a nucleon to a final hadronic state with total isospin
$1/2$ and $3/2$ in the parallel (antiparallel) helicity state $3/2$
($1/2$).  Of all the sum rules that we consider, this one has received
the most attention\footnote{For reviews, see
  Refs.~\cite{Pantforder:1997ii,Drechsel:2000ct}.}, although its basis
for validity is clearly no different from that of the CR sum rule.
The derivation given here in terms of moments of currents parallels
that given in Ref.~\cite{Dicus:1972vp}. Current-algebra derivations
that follow from consideration of the currents themselves rather than their moments are given in
Refs.~\cite{Kawarabayashi:1966zz,Dicus:1973kx,Dicus:1973ae}. These derivations require
consideration of the off-forward Compton amplitude.

%%%%%%%%%%%%%%%%%%%%%%%%%%%%%%%%%%%%%%%%%%%%%%%%%%%%%%%%%%%%%%%%%%%%%%%%%%%%%%%%%%%%%%%%%%%%%%%%%%%%%%%%%%%%%%%%%%%%%%%%%%%%%%%%%%%%%%%%
\subsection{Axial-Vector I}
\label{av}

%%%%%%%%%%%%%%%%%%%%%%%%%%%%%%%%%%%%%%%%%%%%%%%%%%%%%%%%%%%%%%%%%%%%%%%%%%%%%%%%%%%%%%%%%%%%%%%%%%%%%%%%%%%%%%%%%%%%%%%%%%%%%%%%%%%%%%%%
\subsubsection*{General case}
\label{avgc}

\noindent Now we will consider the simplest non-trivial commutators of mixed axial-vector type. As the associated scattering processes are 
inherently non-diagonal, the sum rules will be for the absorptive parts of scattering amplitudes.
Consider first the matrix element between hadronic momentum states of the commutator, Eq.~(\ref{eq:npQDD}):
\BA
\langle\, h'\,,\,\lambda'\,;\,p'\, | {[\, {\tilde Q}_{5\alpha}(x^+) \, ,\, {\tilde {d}}^r_\beta(x^+)\, ]}|\, h \,,\,\lambda\,;\,p\,\rangle  & = &  i\,\epsilon_{\alpha\beta\gamma} \, \langle\, h'\,,\,\lambda'\,;\,p'\, |{\tilde d}^{r}_{5\gamma}(x^+)|\, h \,,\,\lambda\,;\,p\,\rangle \ .
\label{eq:MEphoto}
\EA 
As the axial charge is a scalar and the dipole operator changes the helicity of a state by one unit, this 
commutator decomposes to circular harmonics $\ell=\pm 1$. Therefore, the sum rule
necessarily involves helicity flip and $\Delta\lambda =\ell= \pm 1$.
Following the now familiar procedure, the Born contribution to the LHS is:
\BA
&&{\hspace{-0.5in}}\frac{1}{2p^+} (2\pi)^3 \delta(\,{q^+}\,)
\lbrack\, i\partial^r_{{\bf p}^\prime}\,\delta^2(\,{\bf q}_\perp\,)\,\rbrack \nonumber \\
&&{\hspace{-0.5in}}\times\, \sum_{h''}\Big\lbrack
\langle\, h'\,,\,\lambda'\,;\,p'\,|\, {\tilde J}_{5\alpha}^+(0)  \, |\, h'' \,,\,\lambda''\,;\,p'\,\rangle 
\langle\, h''\,,\,\lambda''\,;\,p'\,|\, {\tilde J}_\beta^+(0)  \, |\, h \,,\,\lambda\,;\,p\,\rangle \nonumber \\
&&\ -\  \langle\, h'\,,\,\lambda'\,;\,p'\,|\, {\tilde J}_{\beta}^+(0)  \, |\, h'' \,,\,\lambda''\,;\,p\,\rangle 
\langle\, h''\,,\,\lambda''\,;\,p\,|\, {\tilde J}_{5\alpha}^+(0)  \, |\, h \,,\,\lambda\,;\,p\,\rangle  \Big\rbrack
\label{eq:born2} 
\EA
and the RHS is 
\BA
(2\pi)^3 \delta(\,{q^+}\,)\lbrack\, i\partial^r_{{\bf p}^\prime}\,\delta^2(\,{\bf q}_\perp \,)\,\rbrack 
i\,\epsilon_{\alpha\beta\gamma} \langle\, h'\,,\,\lambda'\,;\,p'\,|\, {\tilde J}_{5\gamma}^+ (0)  \, |\, h \,,\,\lambda\,;\,p\,\rangle \ .
\label{eq:varhs} 
\EA
The continuum contribution is:
\BA{\hspace{-0.5in}}
(2\pi)^3 \delta(\,q^+\,)\,\delta^2(\,{\bf q}_\perp \,)
\langle h'\,,\,\lambda' |(2\pi)^3
\frac{{\tilde {\cal D}}_{5\alpha}(0)\delta\left(P^+-{p^+}\right)\delta^2\left({\bf P}_\perp-{{\bf p}_\perp}\right)    {\tilde J}_{\beta}^r(0)}
{\left(P^-\,-\,p^{\prime-}\right)\left(P^-\,-\,p^-\right)}| h \,,\,\lambda\rangle  -{\it c.t.}
\label{eq:MEnpRHS4}
\EA
and therefore here we define
\BA w_{\lambda'\lambda;\alpha\beta}^{h'h;r}(p,p',q)& \equiv&
\langle\, h'\,,\,\lambda'\,;\,p'\, |\,(2\pi)^3 \, {\tilde {\cal D}}_{5\alpha}(0) 
\delta^4\left({q}\,+\,{p}\,-{P}\right){\tilde J}_{\beta}^r(0)\, |\, h\,,\,\lambda\,;\,p\, \rangle \ -\ \nonumber \\
&&\langle\, h'\,,\,\lambda'\,;\,p'\, |\,(2\pi)^3 \, {\tilde J}_{\beta}^r(0) 
\delta^4\left({q}\,+\,{p}\,-{P}\right){\tilde {\cal D}}_{5\alpha}(0)\, |\, h\,,\,\lambda\,;\,p\, \rangle \ .
\label{eq:wgamq5defined} 
\EA
And finally the general form of the structure-function sum rule is
\BA
&&\int_{{\bar\nu}_T}^\infty\frac{d{\bar\nu}}{{\bar\nu}\left(2{\bar\nu}+M_h^2-M_{h'}^2\right)}{w}_{\lambda'\lambda;\alpha\beta}^{h'h;r}(p,p',q)|_{q^+={\bf q}_\perp=0} \nonumber \\
&&+\frac{1}{(2p^+)^2}\Big\lbrack\,\sum_{h''}  
\langle\, h'\,,\,\lambda'\,;\,p'\,|\, {\tilde J}_{5\alpha}^+(0)  \, |\, h'' \,,\,\lambda''\,;\,p'\,\rangle 
\big\lbrack i\partial^r_{\bf p}  \langle\, h''\,,\,\lambda''\,;\,p'\,|\, {\tilde J}_\beta^+(0)  \, |\, h \,,\,\lambda\,;\,p\,\rangle  \,\big\rbrack |_{q^+={\bf q}_\perp=0}\nonumber \\
&&- 
{\it c.t.}\Big\rbrack\ =\  
\frac{1}{2p^+}i\epsilon_{\alpha\beta\gamma} \left( i\partial^r_{\bf p}\langle\, h'\,,\,\lambda'\,;\,p'\,|\, {\tilde J}_{5\gamma}^+(0)  \, |\, h \,,\,\lambda\,;\,p\,\rangle \right) |_{q^+={\bf q}_\perp=0} \ .
\label{eq:VAsumrulegeneral}
\EA

%%%%%%%%%%%%%%%%%%%%%%%%%%%%%%%%%%%%%%%%%%%%%%%%%%%%%%%%%%%%%%%%%%%%%%%%%%%%%%%%%%%%%%%%%%%%%%%%%%%%%%%%%%%%%%%%%%%%%%%%%%%%%%%%%%%%%%%%
\subsubsection*{Nucleon sum rule}
\label{avnsr}

\noindent Using the photoproduction conventions of Appendix~\ref{sec:ScattConv}, the nucleon sum rule can be written as
\BA
{\hspace{-0.5in}}&&\int_{-\infty}^{\infty}\frac{d{\bar\nu}}{{\bar\nu}^2}   {\cal C}_{\alpha\beta}^{r}(p,p',q) |_{q^+={\bf q}_\perp=0} \ =\ -2i \xi^T\,
\Big\lbrace {G}^A_{\alpha}(0),\frac{{F}^V_{2\beta}(0)}{2m_N}\Big\rbrace\xi\, \bit{\bar e}^r_\perp
\label{eq:VAsumrulenucleon}
\EA
where $\bit{\bar e}^r_\perp$  is defined in Appendix~\ref{NNPS}.  This then gives 
\BA
\boxed{ g_A \frac{\kappa_V}{2m_N}\ =\ \frac{4F_\pi}{e\pi}\int_{{\bar\nu}_T}^\infty \frac{d{\bar\nu}}{\bar\nu} {\rm Im}\; A_1^{+}({\bar\nu},0)\ .} 
\label{eq:AGsr} 
\EA
This sum rule was first written down by Fubini, Furlan and Rossetti (FFR1)~\cite{Fubini:1965,Adler:1966gd,deAlfaro:1973zz} who considered
the matrix elements of the current commutators at infinite momentum. The more general sum rule which they derived reduces to Eq.~(\ref{eq:AGsr})
in the forward limit. Recent analyses which include chiral corrections using $\chi$PT are given in Refs.~\cite{Arndt:1995ic,Pasquini:2004nq,Bernard:2005dj}. 
As is the case with the GDH sum rule, this sum rule can also be expressed for the isoscalar magnetic moment by choosing the appropriate currents in the derivation. 
There are also equivalent sum rules where weak axial currents couple to the structure functions and constrain neutrino-nucleon interactions~\cite{Gorchtein:2015qha}.

%%%%%%%%%%%%%%%%%%%%%%%%%%%%%%%%%%%%%%%%%%%%%%%%%%%%%%%%%%%%%%%%%%%%%%%%%%%%%%%%%%%%%%%%%%%%%%%%%%%%%%%%%%%%%%%%%%%%%%%%%%%%%%%%%%%%%%%%
\subsection{Axial-Vector II}
\label{av2}

%%%%%%%%%%%%%%%%%%%%%%%%%%%%%%%%%%%%%%%%%%%%%%%%%%%%%%%%%%%%%%%%%%%%%%%%%%%%%%%%%%%%%%%%%%%%%%%%%%%%%%%%%%%%%%%%%%%%%%%%%%%%%%%%%%%%%%%%
\subsubsection*{General case}
\label{av2gc}

\noindent Here we will adopt off-forward kinematics using the electroproduction conventions of Appendix~\ref{sec:ScattConv}. 
Consider the matrix element between hadronic momentum states of the commutator, Eq.~(\ref{eq:npQrr}):
\BA
\langle\, h'\,,\,\lambda'\,;\,p'\, | {[\, {\tilde Q}_{5\alpha}(x^+) \, ,\, {\tilde {r}}^{rs}_\beta(x^+)\, ]}|\, h \,,\,\lambda\,;\,p\,\rangle  & = &  i\,\epsilon_{\alpha\beta\gamma} \, \langle\, h'\,,\,\lambda'\,;\,p'\, |{\tilde r}^{rs}_{5\gamma}(x^+)|\, h \,,\,\lambda\,;\,p\,\rangle \ .
\label{eq:MEphoto2}
\EA 
As in the vector-vector case, non-trivial sum rules are expected for $\Delta\lambda =\ell =0,\pm 2$. The Born contribution to the LHS is:
\BA
&&{\hspace{-0.5in}}\frac{1}{2p^+} (2\pi)^3 \delta(\,{q^+}\,)\lbrack\, i\partial^r_{\bf p}i\partial^s_{\bf p}\,\delta^2(\,{\bf q}_\perp\,)\,\rbrack  \nonumber \\
&&{\hspace{-0.5in}}\times\,\Big\lbrack \sum_{h''}
\langle\, h'\,,\,\lambda'\,;\,p'\,|\, {\tilde J}_{5\alpha}^+(0)  \, |\, h'' \,,\,\lambda''\,;\,p'\,\rangle 
\langle\, h''\,,\,\lambda''\,;\,p'\,|\, {\tilde J}_\beta^+(0)  \, |\, h \,,\,\lambda\,;\,p\,\rangle 
- {\it c.t.}\Big\rbrack
\label{eq:born3} 
\EA
while the RHS is
\BA
(2\pi)^3 \delta(\,{q^+}\,)\lbrack\, i\partial^r_{\bf p}i\partial^s_{\bf p}\,\delta^2(\,{\bf q}_\perp\,)\,\rbrack 
i\,\epsilon_{\alpha\beta\gamma}\langle\, h'\,,\,\lambda'\,;\,p'\,|\, {\tilde J}_{5\gamma}^+(0)  \, |\, h \,,\,\lambda\,;\,p\,\rangle \ .
\label{eq:va2rhs} 
\EA
And the general structure-function sum rule is:
\BA
&&\int_{{\bar\nu}_T}^{\infty}\frac{d{\bar\nu}}{{{\bar\nu}}\left(2{\bar\nu}+M_h^2-M_{h'}^2\right)}\Big\lbrack \,
i\partial^r_{{\bf p}^\prime}{w}_{\lambda'\lambda;\alpha\beta}^{h'h;s}(p,p',q) +
i\partial^s_{{\bf p}^\prime}{w}_{\lambda'\lambda;\alpha\beta}^{h'h;r}(p,p',q) \,
\Big\rbrack |_{q^+={\bf q}_\perp=0}  \nonumber \\
&&-\frac{1}{(2p^+)^2}\Big\lbrack\,\sum_{h''}  
\langle\, h'\,,\,\lambda'\,;\,p'\,|\, {\tilde J}_{5\alpha}^+(0)  \, |\, h \,,\,\lambda''\,;\,p'\,\rangle 
\left(i\partial^r_{\bf p}i\partial^s_{\bf p}  \langle\, h''\,,\,\lambda''\,;\,p'\,|\, {\tilde J}_\beta^+(0)  \, |\, h \,,\,\lambda\,;\,p\,\rangle  \,\rbrack \right)|_{q^+={\bf q}_\perp=0}\nonumber \\
&&- {\it c.t.}\Big\rbrack
\ =\  
-\frac{1}{2p^+}i\epsilon_{\alpha\beta\gamma} \left( i\partial^r_{\bf p}i\partial^s_{\bf p} \langle\, h'\,,\,\lambda'\,;\,p'\,|\, {\tilde J}_{5\gamma}^+(0)  \, |\, h \,,\,\lambda\,;\,p\,\rangle \right) |_{q^+={\bf q}_\perp=0} \ .
\label{eq:VA2sumrulegeneral}
\EA

%%%%%%%%%%%%%%%%%%%%%%%%%%%%%%%%%%%%%%%%%%%%%%%%%%%%%%%%%%%%%%%%%%%%%%%%%%%%%%%%%%%%%%%%%%%%%%%%%%%%%%%%%%%%%%%%%%%%%%%%%%%%%%%%%%%%%%%%
\subsubsection*{Nucleon sum rule}
\label{av2nsr}

\noindent As the nucleon is a spin one-half object, there is no spin flip and therefore the sum rule is proportional to ${\delta^{rs}}$. The need for off-forward
kinematics becomes clear as the sum rule is for an electroproduction amplitude. The helicity label is omitted and the nucleon sum rule is:
\BA
{\hspace{-0.5in}}&&-\frac{1}{4}\int_{-\infty}^{\infty}\frac{d{\bar\nu}}{{\bar\nu}^2} i\partial^r_{{\bf p}^\prime}{\cal C}_{\alpha\beta}^{r}(p,p',q) |_{q^+={\bf q}_\perp=0} 
\ +\  \xi^T\, \lbrack {G}^A_{\alpha}(0),{{F}^{\prime V}_{1\beta}(0)}\rbrack\xi\, 
\ =\ i\epsilon_{\alpha\beta\gamma} \xi^T\, {G}^{\prime A}_{\gamma}(0)\xi\ .
\label{eq:VA2sumrulenucleon}
\EA
Matching to the electroproduction amplitudes in Appendix~\ref{sec:ScattConv}, only a single amplitude contributes,
\BA
\boxed{ \frac{g_A}{6}\left(\;\vev{r^2}^V_1\; -\; \vev{r^2}^A\;\right) \ =\ \frac{4F_\pi}{e\pi}\int_{{\bar\nu}_T}^\infty \frac{d\bar\nu}{\bar\nu} {\rm Im}\; A_6^{-}({\bar\nu},0)\ .} 
\label{eq:aradiussr} 
\EA 
This is a second sum rule first found by Fubini, Furlan and Rossetti
(FFR2)~\cite{Fubini:1965,Adler:1966gd,deAlfaro:1973zz}.  A recent
analysis of this remarkable sum rule that constrains the nucleon axial
radius is given in Ref.~\cite{Pasquini:2007fw,Drechsel:2007gz}, which
incorporate chiral corrections. The sum rule compares very favorably
with experiment and is highly non-trivial as substantial cancellations
take place among contributions from distinct regions of the integral.

\section{Symmetries of the S-matrix}
\label{sec:Ssymms}

\subsection{Regge model expectations}
\label{sec:Reggeme}

\noindent The sum rules found above imply that the imaginary parts of
the relevant forward scattering amplitudes vanish in the Regge limit.
Indeed, as described in the introduction, it is well known that all of
the sum rules can be derived by writing down an appropriate
unsubtracted dispersion relation and assuming the asymptotic behavior
suggested by the Regge model.  Specializing to nucleon targets, Regge
lore suggests the following Regge-limit behavior of the amplitudes for
which sum rules have been derived:
\BA
{{\cal T}_{[\alpha\beta]}({\bar\nu},0)}\ \mapright{{\bar\nu}\rightarrow\infty} \ {\bar\nu}^{\alpha_\rho(0)}
\label{eq:TRegge} 
\EA
for pion scattering, 
\BA
&&{T}_{1[\alpha\beta]}({\bar\nu},0) \mapright{{\bar\nu}\rightarrow\infty} \ {\bar\nu}^{\alpha_\rho(0)-2} \ ;\\
&&{T}_{3(\alpha\beta)}({\bar\nu},0) \mapright{{\bar\nu}\rightarrow\infty} \ {\bar\nu}^{{\tilde\alpha}_P(0)-1} \ ;\\
&&{T}_{4(\alpha\beta)}({\bar\nu},0) \mapright{{\bar\nu}\rightarrow\infty} \ {\bar\nu}^{{\tilde\alpha}_P(0)-2}
\label{eq:WRegge} 
\EA
for Compton scattering on the nucleon and 
\BA
&&{\cal A}_{1(\alpha\beta)]}({\bar\nu},0) \mapright{{\bar\nu}\rightarrow\infty} \ {\bar\nu}^{\alpha_\omega (0)-1} \ ;\\
&&{\cal A}_{6[\alpha\beta]}({\bar\nu},0) \mapright{{\bar\nu}\rightarrow\infty} \ {\bar\nu}^{{\alpha}_{a_1}(0)-1}
\label{eq:ARegge} 
\EA 
for pion photo/electroproduction on the nucleon. Here, $\alpha_M(t)$ (with $M=\rho, \omega, a_1$)
and ${\tilde\alpha}_P(t)$ are the leading odd- and even-signature Regge trajectories of definite G-parity, respectively~\cite{Dicus:1971uk,deAlfaro:1973zz},
corresponding to the appropriate quantum numbers exchanged in the t-channel of the
process~\cite{Collins:1977jy}. Suffice it to say that in all six
cases, the intercepts are expected to be such that unsubtracted
dispersion relations can be written down.  Therefore, the sum rules
derived above from QCD current algebra are consistent with Regge model
expectations, and it is of interest to express the constraints placed
by QCD directly on the Regge-limit amplitudes.

\subsection{Dispersive representation}
\label{sec:dispthe}

\noindent Consider a crossing-even forward scattering amplitude, ${\mathbf{t}}({\bar{\nu}})$, whose
asymptotic behavior is {\it a priori} constrained by the Froissart-Martin bound, which requires  
${\rm Im}\,{\mathbf{t}}({\bar\nu})< {\bar\nu} \ln^2{\bar\nu}$ at large ${\bar\nu}$~\cite{Froissart:1961ux,Martin:1962rt,Hohler:1983}.
This scattering amplitude therefore satisfies a subtracted dispersion relation 
\BA
{\rm Re}\,{\mathbf{t}}({\bar\nu})-{\rm Re}\,{\mathbf{t}}({\bar\nu}_s)\ = \ \frac{2}{\pi}({\bar\nu}^2-{\bar\nu}_s^2)
P\int_{{\bar\nu}_T}^\infty  \frac{{\rm Im}\,{\mathbf{t}}({\bar\nu}'){{\bar\nu}'}d{\bar\nu}'}{({\bar\nu}^{\prime 2}-{\bar\nu}^2)({\bar\nu}^{\prime 2}-{\bar\nu}_s^2)} \ ,
\label{eq:tsubdispersion}
\EA
where $P$ denotes principal value.
At ${\bar\nu}=0$ the scattering amplitude is real and given by
\BA
{\mathbf{t}}(0)\ =\ {\rm Re}\,{\mathbf{t}}({\bar\nu}_s)\ - \ 
\frac{2{\bar\nu}_s^2}{\pi}P\int_{{\bar\nu}_T}^\infty  \frac{{\rm Im}\,{\mathbf{t}}({\bar\nu}')d{\bar\nu}'}{{\bar\nu}^{\prime}({\bar\nu}^{\prime 2}-{\bar\nu}_s^2)} \ .
\label{eq:tsubdispersionnu0}
\EA
Now if ${\rm Im}\,{\mathbf{t}}({\bar\nu})$ vanishes asymptotically, which has been found for the amplitudes relevant to the five sum rules derived above, 
then the limit ${\bar\nu}_s\rightarrow\infty$ can be taken and one finds
\BA
{\mathbf{t}}(\infty)\ =\ {\mathbf{t}}(0)\ -\ \frac{2}{\pi}\int_{{\bar\nu}_T}^\infty  \frac{d{\bar\nu}}{{\bar\nu}} {\rm Im}\,{\mathbf{t}}({\bar\nu}) \ ,
\label{eq:tsubdispersionnu0infty}
\EA
where ${\mathbf{t}}(\infty)$ is the Regge limit amplitude. This is the basic dispersion relation which underlies all of the sum rules that have
been derived above.

\subsection{Regge limit amplitudes}
\label{sec:Reggedisc}

\noindent The nucleon sum rules can now be matched to dispersion theory in order to give explicit expressions for the relevant Regge-limit amplitudes.
This is simply a question of expressing the derivations of Section~\ref{msr} in a slightly different manner. Some detail will be given for the axial-axial case.
Using the results of Section~\ref{sec:aagc} gives
\BA
&&{\hspace{-0.43in}}{\langle\, N;p'\, |\left( i\epsilon_{\alpha\beta\gamma}{\tilde {Q}}_\gamma - {[\, {\tilde {Q}}_{5\alpha}(x^+) \, ,\, {\tilde {Q}}_{5\beta}(x^+)\, ]} \right)|\, N ;p\,\rangle}\ =\ \nonumber  \\
&&{\hspace{-0.3in}}(2\pi)^3\,2\,p^+\,\delta(\,q^+\,)\,\delta^2(\,{\bf q}_\perp \,)\Bigg\lbrack i\,\epsilon_{\alpha\beta\gamma}\,\lbrack\, T_\gamma \,\rbrack_{N}\left(1-g_A^2\right) \ -\ 
\frac{F_\pi^2}{\pi}\int_0^\infty\frac{d{\bar\nu}}{{{\bar\nu}^2}}\,  {\rm Im}\; {\cal T}_{[\alpha\beta]}({\bar\nu},0) \Bigg\rbrack .
\label{eq:AWreggelpre} 
\EA
Now taking ${\mathbf{t}}({\bar\nu})\rightarrow {{\cal T}_{[\alpha\beta]}({\bar\nu},0)}/{\bar\nu}$ and matching Eq.~(\ref{eq:AWreggelpre}) and Eq.~(\ref{eq:tsubdispersionnu0infty}) directly gives the
Regge-limit value of the amplitude\footnote{In addition one gets the correct (chiral-limit) low-energy theorem: $D^-(\nu,0)/\nu|_{\nu=0}=(1-g_A^2)/2F_\pi^2$.}
\BA 
{\hspace{-0.25in}}\boxed{\frac{{\cal T}_{[\alpha\beta]}({\bar\nu},0)}{\bar\nu}|\mapwrong{{\bar\nu}\rightarrow\infty}\quad \frac{2}{F_\pi^2}
\int \frac{d k^+ d^2 \bit{k}_\perp}{2 k^+ (2 \pi)^3}
\langle\, N;k'\, |\left( i\epsilon_{\alpha\beta\gamma}{\tilde {Q}}_\gamma - {[\, {\tilde {Q}}_{5\alpha}(x^+) \, ,\, {\tilde {Q}}_{5\beta}(x^+)\, ]} \right)|\, N ;k\,\rangle .}
\label{eq:AWreggel} 
\EA  
The AW sum rule then follows directly from the axial-vector charge algebra.
If one evaluates this expression in the single-particle approximation using Eqs.~(\ref{eq:isodef}) and (\ref{eq:HME40}), one recovers the result found
originally by Weinberg~\cite{Weinberg:1969hw} who derived the pion-hadron scattering amplitude from the most general chiral Lagrangian and extracted
the leading term in the high-energy expansion. This is gratifying as it demonstrates that these results are independent of the choice of quantization
surface, as of course must be the case.

The Regge-limit Compton amplitudes are derived in similar fashion, giving
\BA
{\hspace{-0.25in}}\boxed{{\bar\nu}\;{T}_{1[\alpha\beta]}({\bar\nu},0) |\mapwrong{{\bar\nu}\rightarrow\infty}\!-e^2{\delta_{rs}}\!\!
\int \frac{d k^+ d^2 \bit{k}_\perp}{2 k^+ (2 \pi)^3}
\langle\, N;k' |\!\left(  i\epsilon_{\alpha\beta\gamma}{\tilde {r}}^{rs}_\gamma(x^+) - {[ {\tilde {d}}^r_\alpha(x^+)  , {\tilde {d}}^s_\beta(x^+) ]}        \right)\!| N ;k\,\rangle}
\label{eq:CRreggel} 
\EA
and
\BA
{\hspace{-0.25in}}\boxed{\!\big\lbrack {T}_{3(\alpha\beta)}({\bar\nu},0) +{\bar\nu} {T}_{4(\alpha\beta)}({\bar\nu},0) \big\rbrack \mapwrong{{\bar\nu}\rightarrow\infty}
\! -\frac{ie^2}{2}\epsilon^{rs}
\!\!\!\int \frac{d k^+ d^2 \bit{k}_\perp}{2 k^+ (2 \pi)^3}
\langle\, N;k'\, | {[\, {\tilde {d}}^r_\alpha(x^+) \, ,\, {\tilde {d}}^s_\beta(x^+)\, ]}|\, N ;k\,\rangle.}
\label{eq:GDHreggel} 
\EA
The second result is homogeneous as the omitted term in the moment
algebra is antisymmetric in the isospin indices.  Hence the CR and the
GDH sum rules follow from the current algebra on the transverse plane
and in particular, Eq.~(\ref{eq:LCalga2d}). It is important to stress
that these sum rules are consequences of more than the symmetry
algebra obeyed by the charges; they rely on the infinite dimensional
symmetry implied by the current algebra.

The pion photo/electroproduction Regge-limit amplitudes are
\BA
{\hspace{-0.25in}}\boxed{{\cal A}_{1(\alpha\beta)}({\bar\nu},0) |\mapwrong{{\bar\nu}\rightarrow\infty}\;\;\, \frac{ie}{2F_\pi}e^r_\perp
\!\int \frac{d k^+ d^2 \bit{k}_\perp}{2 k^+ (2 \pi)^3}
\langle\, N,\uparrow;k'\, | {[\, {\tilde Q}_{5\alpha}(x^+) \, ,\, {\tilde {d}}^r_\beta(x^+)\, ]}|\, N,\downarrow ;k\,\rangle}
\label{eq:FF1reggel} 
\EA
and
\BA
{\hspace{-0.25in}}\boxed{{\cal A}_{6[\alpha\beta]}({\bar\nu},0) | \mapwrong{{\bar\nu}\rightarrow\infty}\!-\!\frac{e}{4F_\pi}\delta^{rs}
\!\!\!\int\! \frac{d k^+ d^2 \bit{k}_\perp}{2 k^+ (2 \pi)^3}
\langle N;k' |\!\!\left(\! i\epsilon_{\alpha\beta\gamma}{\tilde {r}}^{rs}_{5\gamma}(x^+) - {[ {\tilde Q}_{5\alpha}(x^+)  , {\tilde {r}}^{rs}_\beta(x^+) ]} \!\right)\!\!| N ;k\rangle .}
\label{eq:FF2Rreggel} 
\EA 
The first result is homogeneous as the omitted term is antisymmetric in the isospin indices.
From these Regge-limit amplitudes, the FFR1 and FFR2 sum rules follow, respectively,
from the infinite dimensional symmetry implied by current algebra on the transverse
plane, and in particular, Eq.~(\ref{eq:LCalgb2d}).

Eqs.~(\ref{eq:AWreggel}-\ref{eq:FF2Rreggel}), are the most important
results of this paper as they are explicit expressions of Regge-limit
amplitudes as matrix elements of QCD operators. As pointed out in the
discussion of Section~\ref{NPQCDconC}, if the Regge-limit amplitudes
for Compton scattering and/or photo/electroproduction were found to be
non-vanishing experimentally, that would indicate the presence of
additional terms in the current algebra which involve transverse
gradients\footnote{In the language of the Regge model, this
  corresponds to a fixed pole.}. The existence of such terms would
break the infinite-dimensional chiral symmetry on the transverse plane
and leave only the global symmetry. From the perspective of the closed
algebra of Eqs.~(\ref{eq:fullKMa}-\ref{eq:fullKMc}), if these brackets
are augmented to include central extensions, then the Regge-limit
amplitudes are related to the matrix elements of these extensions.  As
evidenced by the experimental success of the sum rules, these terms
are effectively absent.

While obtained in the chiral limit, the Regge-limit equations are
expected to remain valid as the light-quark mass matrix is turned on,
and therefore, together with dispersion theory, they provide a
starting point for the derivation of sum rules that properly account
for both chiral corrections and effects due to isospin violation (see,
for instance, Ref.~\cite{Klco:2015}).  It is also worth noting that in
order not to violate the Froissart-Martin
bound~\cite{Froissart:1961ux,Martin:1962rt} in the large-$N_c$ limit
of QCD, the algebraic constraints recover the large-$N_c$ consistency
condition for the baryon axial
charges~\cite{Gervais:1983wq,Dashen:1993as,Dashen:1993jt}, and for the
baryon magnetic moment and transition magnetic
moments~\cite{Jenkins:1994md}, as they must\footnote{See also
  Refs.~\cite{Broniowski:1994gm,Wirzba:1993nc}.}.

\subsection{Transverse multipole expansion}
\label{sec:multi}

\noindent As the forward limit probes long-distance scales, the various commutators formed from the transverse moments of the 
vector and axial-vector charge distributions clearly obey a hierarchy of scales on the transverse plane. Consider the following
operator
\BA
{\cal H}_{(5)\alpha}({\bf R}_\perp) \equiv \int d^2{\bf r}_\perp {\tilde F}_{(5)\alpha}({\bf r}_\perp)\,\Phi({\bf r}_\perp+{\bf R}_\perp)\ ,
\label{eq:multi1} 
\EA
where ${\tilde F}_{(5)\alpha}({\bf r}_\perp)$ is the (axial-) vector charge distribution defined in Eq.~(\ref{eq:npchargesBdefined2d})
and $\Phi({\bf r}_\perp)$ is a c-number test function that is slowly varying in the vicinity of the charge distribution. Note that we omit dependence
on the null-plane time. By considering the nucleon matrix
elements of the commutators of Eq.~(\ref{eq:multi1}) with the various transverse moments, one sees the Regge-limit amplitudes emerging 
as the coefficients of the associated multipole expansion. For instance, one finds
\BA
&&{\hspace{-0.45in}}\int \frac{d k^+ d^2 \bit{k}_\perp}{2 k^+ (2 \pi)^3}
\langle\, N;k'\, |\left( [\,{\tilde Q}_{5\alpha}\,,\,{\cal H}_{\beta}({\bf R}_\perp)\,]- i\epsilon_{\alpha\beta\gamma} {\cal H}_{5\gamma}({\bf R}_\perp) \right)|\, N ;k\,\rangle \ =\ \nonumber \\
&& \qquad \frac{F_\pi}{e}\Big\lbrack {\cal A}_{1(\alpha\beta)}({\infty},0)\, i {\bar e}^r_\perp \nabla^r \Phi({\bf R}_\perp)\ +\ 
 {\cal A}_{6[\alpha\beta]}({\infty},0) \,\nabla^2 \Phi({\bf R}_\perp) \ +\  \ldots \ \Big\rbrack \ ,
\label{eq:multi2} 
\EA
where the trivial commutators have been evaluated to zero. This expansion is interesting as it suggests the existence of an effective field theory that can be formulated
entirely on the transverse plane with operators whose coefficients give non-trivial information about high-energy scattering
in QCD.

%%%%%%%%%%%%%%%%%%%%%%%%%%%%%%%%%%%%%%%%%%%%%%%%%%%%%%%%%%%%%%%%%%%%%%%%%%%%%%%%%%
\section{Discussion and Conclusions}
\label{sec:concs}

\noindent This work has considered the forward scattering of vector
and axial-vector currents on hadronic targets. It has long been known
that there is a class of amplitudes whose observed asymptotic behavior
is sufficiently soft (softer than the unitarity bound) as to allow the
use of unsubtracted dispersion relations to relate the integral over
the imaginary part to the amplitude at a low-energy point where it is
known from soft-pion or soft-photon theorems. This behavior of forward
scattering amplitudes is usually justified using Regge pole theory,
which is a well-motivated and highly-successful phenomenological
description whose origins in QCD are unclear.  In this
paper, it has been shown that there is a class of scattering
amplitudes whose Regge limit values can be calculated exactly, and
expressed as matrix elements of null-plane QCD operators, whose
behavior is constrained by QCD current algebra on the transverse plane.
Specifically, the main conclusions of this paper are:

\vskip0.2in
\noindent$\bullet$ Quantizing QCD on light-like hyperplanes leads to
light-cone current algebras that are useful for deriving sum rules for
processes that scatter currents from arbitrary hadronic targets. In
particular, the infinite-dimensional algebra satisfied by null-plane
axial-vector and vector charge distributions leads to sum rules that are
profitably derived by considering the Lie brackets satisfied by the
moments of the charge distributions. By considering the simplest
relations, a system of five sum rules was derived, which constrain the
scattering of vector and axial currents from arbitrary hadronic
targets in accord with the allowed symmetries of QCD.  For the case of
a nucleon target in the initial and final state, interacting with pions and
isovector photons, these sum rules reduce to the Adler-Weisberger (Eq.~(\ref{eq:AWsumrule}), 
Cabbibo-Radicatti (Eq.~(\ref{eq:CRSR}), Gerasimov-Drell-Hearn (Eq.~(\ref{eq:GDHSR}), 
and Fubini-Furlan-Rossetti sum rules (Eq.~(\ref{eq:AGsr} and Eq.~(\ref{eq:aradiussr})).
One interesting implication of this result is that these well-known sum rules
are seen to share the same origin.

\vskip0.2in
\noindent$\bullet$ The most general way of expressing the sum rules
derived in this paper is as statements about the relevant S-matrix
elements in the Regge limit, Eqs.~(\ref{eq:AWreggel}-\ref{eq:FF2Rreggel}). In all cases considered, these
constraints derived from QCD are consistent with Regge lore. As the
algebraic constraints, which are derived in the chiral limit, are not
expected to change form in the presence of explicit breaking of chiral
symmetry, they serve ---when combined with dispersion theory--- as the
starting point for a derivation of a physical formulation of the sum rules
(involving pions) which incorporate the correct quark mass corrections
rigorously.

\vskip0.2in
\noindent$\bullet$ The moment sum rules illustrate the fundamental
importance that algebraic chiral symmetry places on the hadronic world
via asymptotic constraints on S-matrix elements.  It is clear from the
derivation of the sum rules that the algebraic chiral symmetry knows
about dynamical chiral symmetry and vice-versa. For instance, from the
Lie bracket of axial charges, one can match onto dispersion theory and
obtain the scattering amplitude at threshold (see footnote 10), that
is, the soft-pion theorem, which is computed using leading-order
$\chi$PT in the chiral limit. This is not surprising as the current
algebra serves as a chiral Ward identity. Alternatively, one can start
in $\chi$PT in the chiral limit with non-linear realization of the
chiral symmetry, but include all allowed operators in the calculation
of the forward amplitude. One then sees that the Lie bracket satisfied
by the axial charges is present in the leading term in a high-energy
expansion of the forward amplitude.  

\vskip0.2in
\noindent$\bullet$ A striking aspect of the results derived in this
paper is the two-dimensional nature of the constraints which arise
from the infinite-dimensional chiral symmetry that exists on the
transverse plane. The two-dimensional nature of the sum rules is
rather natural from the perspective of null-plane quantization and has
an interesting and intuitive partonic interpretation using
renormalization group arguments (see, for instance,
Ref.~\cite{Casher:1974xd}). Here it should be emphasized that from the
perspective of QCD degrees of freedom, it is natural to expect that
high-energy scattering, and particularly the Regge limit, is governed
by the dynamics of partons on the transverse
plane~\cite{Lipatov:1994xy,Belitsky:2004cz}.  
\vskip 0.2in

We have focused specifically on those aspects of the Regge limit
that appear to be constrained solely by symmetry, which arise, for instance, in the
differences of cross-sections. The total cross-sections, governed by
the Pomeron in the Regge model language, are also constrained by the
current algebra. However, these constraints do not have the simple
symmetry interpretation that we have seen here as they involve light-ray
operators (light-like correlations). They are therefore substantially
more complex and will be considered elsewhere.

\vskip 0.2in

\noindent We thank A.~Cherman, M.~Hoferichter and M.J.~Savage for valuable
conversations.  The work of SRB was supported in part by
the U.S.  National Science Foundation through continuing grant
PHY1206498 and by the U.S. Department of Energy through Grant Number
DE-SC001347, and work of TJH was supported in part by the
U. S. Department of Energy Office of Science, Office of Nuclear
Physics under Award Number DE-FG02-97ER-41014.

%\vfill\eject

%***************************Appendix*********************************
%

\appendix

\section{Null-plane conventions}\label{npconventions}

\subsection*{Coordinate conventions}
\label{NPQCDconA}

\noindent Here we will introduce basic coordinate conventions that
will be used throughout the paper. Many additional details and
derivations are given in Ref.~\cite{Beane:2013oia} and in the many
reviews on null-plane
quantization~\cite{Lepage:1980fj,Hornbostel:1990ya,Perry:1994kp,Zhang:1994ti,Burkardt:1995ct,Harindranath:1996hq,Brodsky:1997de,Perry:1997uv,Miller:1997cr,Venugopalan:1998zd,Heinzl:2000ht,Miller:2000kv,Diehl:2003ny,Belitsky:2005qn,Mustaki:1994mf,Yamawaki:1998cy,Itakura:2001yt}.

Consider the light-like vectors $n^\mu$ and ${\bar n}^{\mu}$
which satisfy $n^2 = {\bar n}^{2} = 0$ and $n \cdot {\bar n} = 1$. Here we will
choose these vectors such that 
\begin{equation}
n^\mu \equiv \ft{1}{\sqrt{2}} (1, 0, 0, -1) \quad , \quad
{\bar n}^{\mu} \equiv \ft{1}{\sqrt{2}} (1, 0, 0, 1)
\, .
\label{nullvectors}
\end{equation}
In the front-form one chooses the initial quantization surface of the system to be on a light-like plane,
or null-plane, which is a hypersurface of points in Minkowski
space such that $x\cdot n =\tau$. Here $\tau$ is a
constant which plays the role of time. A null-plane is represented as
$\Sigma_n^\tau$.  We will take the initial surface to be the null-plane
$\Sigma_n^0$. A coordinate system adapted to null-planes is then specified by
the four-vector, $x^\mu=(x^+,{x}^1,{x}^2,x^-)=(x^+,{\bf x}_\perp,x^-)$, where
\BA
x^+ \equiv x \cdot n = \ft{1}{\sqrt{2}} (x^0 + x^3) = \tau
\, , \qquad
x^- \equiv x \cdot {\bar n} = \ft{1}{\sqrt{2}} (x^0 - x^3)
\label{eq:LCcoordinates}
\EA are taken as the time variable and ``longitudinal'' lightlike position, respectively.
The remaining spacelike coordinates, ${\bf x}_\perp=( {x}^1, {x}^2)$ provide the ``transverse''
position. Note that the components of a transverse vector do not
have a $\perp$ symbol; e.g. ${\bf x}^2_\perp=x^rx^r$. The metric tensor associated
with this coordinate choice is
\BA
       {\tilde g}^{\mu\nu}\ =\ {\tilde g}_{\mu\nu}\ =\ \begin{pmatrix}0&0&0&1 \cr 
                                                            0&-1&0&0 \cr 
                                                            0&0&-1&0 \cr 
                                                            1&0&0&0 \cr \end{pmatrix} \ ,
\label{eq:LCmetric2}
\EA 
and therefore only the raising and lowering of transverse indices incur a sign change.
Translational invariance implies
\BA
{\cal O}(x) & = & e^{iP\cdot x}\,{\cal O}(0)\, e^{-iP\cdot x} \ .
\label{eq:OLI}
\EA 
It then follows that a general operator ${\cal O}$ with non-trivial null-plane time dependence
satisfies
\BA
{[\, {\cal O}(x) \, ,\, P^-\, ]} & = &  i\,\partial_+ {\cal O}(x) \ .
\label{eq:gentdep}
\EA 
Here $P^-$ is the Hamiltonian that evolves a physical system in
null-plane time. 

\subsection*{Momentum eigenstates}
\label{NPQCDconE}

\noindent We will write a general hadronic momentum eigenstate as:
\BA
|h\, ,\,\lambda \,\rangle \equiv\ |h\, ,\,\lambda\,;\, p^+\, ,\, {\bf p}_\perp\,\rangle \ \equiv \ |h\, ,\,\lambda\,;\, p\,\rangle \ .
\label{eq:HBdefined}
\EA
Note that the label $p$ in the rightmost ket is shorthand for the momentum variables and should
not be confused with the four-vector $p=(p^+,{\bf p}_\perp,p^-)$.
Here the label $h$ includes any additional variables that may be needed to specify the
state of the hadron $h$ when it is at rest, and $\lambda$ is the (total) helicity, the eigenvalue of the
kinematical Poincar\'e generator, ${\cal J}_3$:
\BA
{\cal J}_3\;|h\, ,\,\lambda \,\rangle \ =\ \lambda\;|h\, ,\,\lambda \,\rangle  \, .
\label{eq:J3eigen}
\EA 
In the rest frame of the hadron, ${\cal J}_3=J_3$, the angular momentum operator~\cite{Leutwyler:1977vy}.
The norm of the momentum state is determined by unitarity up to a constant~\cite{Leutwyler:1977vy}, 
\BA
\langle\,p^{\prime+}\,,\, {\bf p}^\prime_\perp\, |\,p^+\,,\, {\bf p}_\perp\,\rangle \ \propto \ 2p^+ \delta(\,p^{\prime +}\,-\,p^+)\;\delta^2(\,{\bf p}^{\prime}_\perp\,-\,{\bf p}_\perp \,)
\label{eq:normmomstate}
\EA
which we choose so that the completeness relation is:
\BA
\sum_{\it \lambda''}\int \frac{d k^+ d^2 \bit{k}_\perp}{2 k^+ (2 \pi)^3}
|\,\lambda''\,;\, k^+\, ,\, {\bf k}_\perp      \,\rangle\, \,\langle\, \lambda''\,;\, k^+\, ,\, {\bf k}_\perp  \, | 
\ =\ {\bf 1} \ .
\label{eq:completeness} 
\EA
Note that ${P}=(P^+,{\bf P}_\perp,P^-)$ is the four-vector energy-momentum operator with 
dispersion relation
\begin{eqnarray}
P^- |h\, ,\,\lambda\,;\, p\,\rangle \ =\ \frac{M^2 + {\bf P}^2_{\perp}}{2P^+}|h\, ,\,\lambda\,;\, p\,\rangle \ =\
\frac{M_h^2 + {\bf p}^2_{\perp}}{2p^+}|h\, ,\,\lambda\,;\, p\,\rangle \ =\ 
p^- |h\, ,\,\lambda\,;\, p\,\rangle \ .
\label{eq:hhpkinematics} 
\end{eqnarray}
The
momentum-state selection rules relevant to QCD symmetries are worked
out in detail in Ref.~\cite{Soper:1972xc}.

\subsection*{Nucleon null-plane spinors}
\label{NNPS}

\noindent We will adopt the conventions of Ref.~\cite{Kogut:1969xa} which makes use of
the Dirac representation of the gamma matrices. What follows is an extension of results from
Ref.~\cite{Belitsky:2005qn}. (See also Ref.~\cite{Brodsky:2006ez}.)
Defining the holomorphic and anti-holomorphic transverse
momenta:
\begin{equation}
p_\perp \equiv p^1 + i p^2
\qquad , \qquad
\bar p_\perp \equiv p^1 - i p^2 \ ,
\end{equation}
and with the total Dirac bispinor normalized as
\begin{equation}
\bar u_{\lambda'} (p) u_{\lambda} (p)
=
2 m_N \, \delta_{\lambda' \lambda}
\, ,
\label{eq:unormal}
\end{equation}
the nucleon spinors with definite null-plane helicity can be written as
\begin{equation}
\label{LightConeSpinors}
u_\uparrow (p)
= \frac{1}{\sqrt[4]{2} \sqrt{p^+}}
\left(
\begin{array}{c}
p^+ + m_N/\sqrt{2} \\
p_\perp/\sqrt{2} \\
p^+ - m_N/\sqrt{2} \\
p_\perp/\sqrt{2}
\end{array}
\right)
\quad , \quad
u_\downarrow (p)
= \frac{1}{\sqrt[4]{2} \sqrt{p^+}}
\left(
\begin{array}{c}
- \bar p_\perp/\sqrt{2} \\
p^+ + m_N/\sqrt{2}        \\
\bar p_\perp/\sqrt{2}   \\
- p^+ + m_N/\sqrt{2}
\end{array}
\right)
\, ,
\end{equation}
where $\textstyle{\uparrow}$ ($\textstyle{\downarrow}$) denotes $\lambda=1/2$ ($-1/2$).
The bilinears that are relevant for this paper are obtained
from the explicit form of the null-plane helicity spinors,
\begin{equation}
\begin{array}{ll}
\!\!
\bar u_\uparrow (p_2) \gamma^+ u_\uparrow (p_1)
=
2 \sqrt{p_1^+ p_2^+}
\, ,
&\quad\!\!\!
\bar u_\downarrow (p_2) \gamma^+ u_\uparrow (p_1)
=
0
\, , \\
\!\!
\bar u_\uparrow (p_2) \sigma^{+ r} u_\uparrow (p_1)
=
0
\, ,
&\quad\!\!\!
\bar u_\downarrow (p_2) \sigma^{+ r} u_\uparrow (p_1)
=
2 i \sqrt{p_1^+ p_2^+} \bit{e}^r_\perp
\, , \\
\!\!
\bar u_\uparrow (p_2) \sigma^{+-} u_\uparrow (p_1)
=
\frac{i m_N (p_2^+ - p_1^+)}{\sqrt{p_1^+ p_2^+}}
\, ,
&\quad\!\!\!
\bar u_\downarrow (p_2) \sigma^{+-} u_\uparrow (p_1)
=
\frac{i (p_1^+ p_{2 \perp} + p_2^+ p_{1 \perp})}{\sqrt{p_1^+ p_2^+}}
, \\
\!\!
\bar u_{\uparrow(\downarrow)} (p_2) \gamma^+ \gamma^5 u_{\uparrow(\downarrow)} (p_1)
=
(-)2 \sqrt{p_1^+ p_2^+}
\, ,
&\quad\!\!\!
\bar u_\downarrow (p_2) \gamma^+ \gamma^5 u_\uparrow (p_1)
=
0
\, , \\
\!\!
\bar u_\uparrow (p_2) \gamma^5 u_\uparrow (p_1)
=
\frac{m_N (p_1^+ - p_2^+)}{\sqrt{p_1^+ p_2^+}}
\, ,
&\quad\!\!\!
\bar u_\downarrow (p_2) \gamma^5 u_\uparrow (p_1)
=
\frac{(p_2^+ p_{1 \perp} - p_1^+ p_{2 \perp})}{\sqrt{p_1^+ p_2^+}}
\, , \\
\!\!
\bar u_\uparrow (p_2) \gamma_5\sigma^{+ r}u_\uparrow (p_1)
=
0
\, ,
&\quad\!\!\!
\bar u_\downarrow (p_2)\gamma_5 \sigma^{+ r} u_\uparrow (p_1)
=
2 i \sqrt{p_1^+ p_2^+} \bit{e}^r_\perp
\, , \\
\!\!
\bar u_\uparrow (p_2) \gamma^r u_\uparrow (p_1)
=
\frac{ (\bit{e}^r_\perp p_1^+ \bar p_{2 \perp} \; +\; \bit{\bar e}^r_\perp p_2^+ p_{1 \perp})}{\sqrt{p_1^+ p_2^+}}
\, ,
&\quad\!\!\!
\bar u_\downarrow (p_2) \gamma^r u_\uparrow (p_1)
=
\frac{\bit{e}^r_\perp m_N (p_1^+ - p_2^+)}{\sqrt{p_1^+ p_2^+}}
\, , \\
\!\!
\bar u_\uparrow (p_2) \gamma_5\gamma^r u_\uparrow (p_1)
=
-\frac{ (\bit{e}^r_\perp p_1^+ \bar p_{2 \perp} \; +\; \bit{\bar e}^r_\perp p_2^+ p_{1 \perp})}{\sqrt{p_1^+ p_2^+}}
\, ,
&\quad\!\!\!
\bar u_\downarrow (p_2) \gamma_5\gamma^r u_\uparrow (p_1)
=
-\frac{\bit{e}^r_\perp m_N (p_1^+ + p_2^+)}{\sqrt{p_1^+ p_2^+}}
\, ,
\end{array}
\label{blinearsNP}
\end{equation}
where the two-dimensional vectors are $\bit{e}^r_\perp = (1, i)$ and $\bit{\bar e}^r_\perp = (1, -i)$ and
satisfy $\bit{\bar e}^r_\perp \bit{e}^s_\perp =\delta^{rs}+i\epsilon^{rs}$.
Note that not all of the relations in Eq.~(\ref{blinearsNP}) are independent.

%\newpage

\section{Nucleon scattering conventions}\label{sec:ScattConv}

\subsection*{Forward pion scattering}%\label{sec:PNSC}

\noindent We use the standard pion-nucleon scattering conventions of
Ref.~\cite{Hohler:1983}.  In particular the energy is expressed 
in terms of the variable $\nu={(s-u)}/{4 m_N}={\bar\nu}/m_N$. As we are working
strictly in the chiral limit, some care is needed in the definition
of the amplitude. The off-shell Green function relevant to pion-nucleon scattering can be written
\BA 
\hspace{-0.25in}{\cal T}_{\alpha\beta}(p',q';p,q) & =&  
i(M_\pi^2-q^2)(M_\pi^2-q^{\prime 2}) \int d^4x\,e^{i {q}\cdot x} \,\langle\, N\,;\,p'\,| {\rm T}\left( {\tilde\pi}_\alpha(x) {\tilde\pi}_\beta(0)\right)  |\, N \,;\,p\,\rangle \ .
\label{eq:piNGF}
\EA
Using Eq.~(\ref{eq:chidivqcd}), and placing the pions on their mass shell (i.e. taking the chiral limit)
gives the pion-nucleon scattering amplitude, ${\cal T}_{\alpha\beta}(\nu, t)$. In the forward direction,
\BA 
\hspace{-0.25in}{\cal T}_{\alpha\beta}(\nu,0)={\cal T}_{\alpha\beta}(p,q) =  \frac{i}{F_\pi^2} 
\int d^4x\,e^{i {q}\cdot x} \,\langle\, N\,;\,p\,| {\rm T}\left( {\tilde {\cal D}}_{5\alpha}(x) {\tilde {\cal D}}_{5\beta}(0)\right)  |\, N \,;\,p\,\rangle \ .
\label{eq:piNSA}
\EA 
The crossing properties of the amplitude follow from the symmetry of the time-ordered product
\BA 
\hspace{-0.25in}{\cal T}_{\alpha\beta}(\nu,0)\ =\ {\cal T}_{\beta\alpha}(-\nu,0) \ .
\label{eq:piNSAcross}
\EA 
The forward amplitude has the standard decomposition
\BA
{\cal T}_{\alpha\beta}(\nu,0)& =&\bar{u}_\lambda(p) D_{\alpha\beta}(\nu, 0) u_\lambda (p) \ ,
\label{eq:sadefs}
\EA
with
\BA
D_{\alpha\beta} & =&  \delta_{\alpha\beta}{D}^+ \ +\ 2i\epsilon_{\alpha\beta\gamma}T_\gamma {D}^- \ .
\label{eq:sadefs2}
\EA
Therefore
\BA
{\cal T}_{[\alpha\beta]}(\nu,0)& =& 2m_N \xi^T\left( 2i\epsilon_{\alpha\beta\gamma}T_\gamma D^-(\nu,0)\right)\xi \ =\ 4 i m_N \epsilon_{\alpha\beta\gamma} \lbrack\, T_\gamma \,\rbrack_{N} D^-(\nu,0) \ ,
\label{eq:sadefs3}
\EA
where Eq.~(\ref{eq:unormal}) has been used. It follows from Eq.~(\ref{eq:piNSAcross}) that
\BA
{D}^{\pm}(\nu,0) \ =\ \pm{D}^{\pm}(-\nu,0) \ .
\label{eq:pincrossing} 
\EA
Defining
\BA 
\hspace{-0.25in}{\cal C}_{\alpha\beta}(p,q) \equiv  \frac{1}{2\pi}
\int d^4x\,e^{i {q}\cdot x} \,\langle\, N\,;\,p\,| [{\tilde {\cal D}}_{5\alpha}(x), {\tilde {\cal D}}_{5\beta}(0)]  |\, N \,;\,p\,\rangle \ ,
\label{eq:piNSAimag}
\EA 
one then finds
\BA 
{\rm Im}\,{\cal T}_{\alpha\beta}(p,q)\ =\ \frac{\pi}{F_\pi^2}{\cal C}_{\alpha\beta}(p,q) \ =\ \frac{\pi}{F_\pi^2} w_{\lambda;\alpha\beta}^{NN}(p,q) \ ,
\label{eq:piNSAimagC}
\EA 
where $\lambda=\pm1/2$. Note that the imaginary part of the scattering amplitude has crossing-symmetry properties that are opposite to that of the scattering amplitude itself
due to the presence of the commutator. 
The optical theorem states
\BA
\mbox{Im}D^\pm(\nu,0)=k\,\sigma^\pm_{tot}(\nu) \ ,
\label{eq:opticalT}
\EA
where $k=\sqrt{\nu^2-M_\pi^2}=\nu$ is the lab momentum of the incoming pion,
and 
\BA
\sigma^\pm_{tot}(\nu) =\oneht \left( \sigma^{\pi^-p}_{tot}(\nu) \ \pm\ \sigma^{\pi^+p}_{tot}(\nu)\right) \ .
\label{eq:csections}
\EA

\subsection*{Forward Compton scattering}%\label{sec:CSSC}

\noindent The forward Compton scattering amplitude is
\BA 
\hspace{-0.25in}{\cal T}^{\mu\nu}_{\alpha\beta}(p,q) & =&  {i}e^2
\int d^4x\,e^{i {q}\cdot x} \,\langle\, N\,;\,p\,| {\rm T}\left( {\tilde J}_\alpha^\mu(x){\tilde J}_\beta^\nu(0) \right)  |\, N \,;\,p\,\rangle \ ,
\label{eq:ComptNSA}
\EA 
and can be decomposed as
\BA
&&{T}_{\alpha\beta}^{\mu\nu}(p,q) \ =\ 
\big\lbrack q^2 p^\mu p^{\nu}-{\bar\nu}\left( p^\mu q^\nu  + p^\nu q^\mu \right)+{\bar\nu}^2 g^{\mu\nu}\,\big\rbrack\, T_{1\alpha\beta}({\bar\nu}, q^2)\ +\ \left( q^\mu q^\nu - g^{\mu\nu}\right) T_{2\alpha\beta}({\bar\nu}, q^2) \nonumber \\
&&\qquad\qquad\qquad\qquad\qquad +\ i\epsilon^{\mu\nu\kappa\delta}s_{\kappa}q_{\delta} T_{3\alpha\beta}({\bar\nu}, q^2) \ +\  \ i q\cdot s \epsilon^{\mu\nu\kappa\delta}p_{\kappa}q_{\delta} T_{4\alpha\beta}({\bar\nu}, q^2) \ ,
\label{eq:Tdefinedinampsnew} 
\EA
where $s^\mu={\bar u}_\lambda (p)\gamma^\mu\gamma_5{u}_\lambda (p)$ is the nucleon four-vector spin. 
The crossing properties of the amplitude follow from the symmetry of the time-ordered product
\BA 
\hspace{-0.25in}{\cal T}^{\mu\nu}_{\alpha\beta}(\nu,0)\ =\ {\cal T}^{\nu\mu}_{\beta\alpha}(-\nu,0) \ .
\label{eq:gamNSAcross}
\EA 
One can further define
\BA
&&{\cal C}^{\mu\nu}_{\alpha\beta}(p,q) \ =\ \frac{1}{2\pi}
\int d^4x\,e^{i {q}\cdot x} \,\langle\, N\,;\,p\,|\, [\,{\tilde J}_\alpha^\mu(x)\,,\, {\tilde J}_\beta^\nu(0)\,]\, |\, N \,;\,p\,\rangle \ ,
\label{eq:cdefinedinamps} 
\EA
where
\BA
&&{\cal C}^{rs}_{\alpha\beta}(p,q) \ =\ w_{\lambda'\lambda;\alpha\beta}^{NN;rs}(p,q)\ .
\label{eq:cdefinedinampsrs} 
\EA
A Lorentz-invariant decomposition is
\BA
&&{C}_{\alpha\beta}^{\mu\nu}(p,q) \ =\ 
\big\lbrack q^2 p^\mu p^{\nu}-{\bar\nu}\left( p^\mu q^\nu  + p^\nu q^\mu \right)+{\bar\nu}^2 g^{\mu\nu}\,\big\rbrack\, W_{1\alpha\beta}({\bar\nu}, q^2)\ +\ \left( q^\mu q^\nu - g^{\mu\nu}\right) W_{2\alpha\beta}({\bar\nu}, q^2) \nonumber \\
&&\qquad\qquad\qquad\qquad\qquad +\ i\epsilon^{\mu\nu\kappa\delta}s_{\kappa}q_{\delta} W_{3\alpha\beta}({\bar\nu}, q^2) \ +\  \ i q\cdot s \epsilon^{\mu\nu\kappa\delta}p_{\kappa}q_{\delta} W_{4\alpha\beta}({\bar\nu}, q^2) \ .
\label{eq:cdefinedinampsnew} 
\EA
It is straightforward to find
\BA
{\rm Im}\;{T}_{i\alpha\beta}({\bar\nu},q^2) \ =\ \pi {e^2} {W}_{i\alpha\beta}({\bar\nu},q^2) \ .
\label{eq:cdefinedinamps2} 
\EA
Again the imaginary part has opposite crossing-symmetry properties to the amplitude itself.
The crossing properties that are relevant for this paper are
\BA
W_{1[\alpha\beta]}({\bar\nu},q^2) & =&  W_{1[\alpha\beta]}(-{\bar\nu},q^2) \ ; \\
W_{3(\alpha\beta)}({\bar\nu},q^2) & =&  -W_{3(\alpha\beta)}(-{\bar\nu},q^2) \ ; \\
W_{4(\alpha\beta)}({\bar\nu},q^2) & =&  W_{4(\alpha\beta)}(-{\bar\nu},q^2) \ .
\label{eq:comptoncrossing} 
\EA
Defining
\BA
W_{1[\alpha\beta]}({\bar\nu},q^2) & =&  2 i \epsilon_{\alpha\beta\gamma} \lbrack\, T_\gamma \,\rbrack_{N} W_{1}({\bar\nu},q^2) \ ; \\
W_{3(\alpha\beta)}({\bar\nu},q^2) & =&  \delta_{\alpha\beta} W_{3}({\bar\nu},q^2) \ ; \\
W_{4(\alpha\beta)}({\bar\nu},q^2) & =&  \delta_{\alpha\beta} W_{4}({\bar\nu},q^2)  \ ,
\label{eq:comptoncrossingdefinew1} 
\EA
one has the optical theorems~\cite{deAlfaro:1973zz}
\BA
\pi e^2\;{\bar\nu }\;W_{1}({\bar\nu},0) & =& \oneht \left( \sigma^{\gamma^-p}_{tot}({\bar\nu}) \ -\ \sigma^{\gamma^+p}_{tot}({\bar\nu})\right) \ = \ -\left(\, 2\sigma^V_{1/2}({\bar\nu})\ -\ \sigma^V_{3/2}({\bar\nu})\, \right)
\label{eq:W1opttheorem} 
\EA
where in the last step an isospin rotation has been performed, and 
\BA
\pi e^2\;{\bar\nu }\; \left( {W}_{3}({\bar\nu},0)\ +\ {\bar\nu} {W}_{4}({\bar\nu},0)\right) & =& \oneht \left(\, \sigma^V_{P}({\bar\nu})\ -\ \sigma^V_{A}({\bar\nu})\, \right) \ .
\label{eq:W1opttheorem2} 
\EA
The various cross-sections are defined in the text.

\subsection*{Photo/electroproduction}%\label{sec:NPSC}

\noindent The electroproduction amplitude is 
\BA
{\cal T}_{\alpha\beta}^{\mu}(p,p',q)& =& 
\frac{ie}{F_\pi} 
\int d^4x\,e^{i {q}\cdot x} \,\langle\, N\,;\,p'\,| {\rm T}\left( {\tilde {\cal D}}_{5\alpha}(x) {\tilde J}^\mu_{\beta}(0)\right)  |\, N \,;\,p\,\rangle \ ,
\label{eq:electroamp}
\EA 
and can be decomposed into Lorentz invariant amplitudes as~\cite{Bernard:1995dp}
\BA
{\cal T}_{\alpha\beta}^{\mu}(p,p',q)& =& 
i{\bar u}_{\lambda'}(p')\,\gamma_5\;\sum_{i=1}^{6}\;{\vartheta}_i^\mu\;{A}^i_{\alpha\beta}({\bar\nu},t,q^2)\;{u}_\lambda (p) \ 
\label{eq:photoTdefined} 
\EA
where 
\BA
&&\vartheta_1^\mu = {1\over 2} (\gamma^\mu \qs - \qs \gamma^\mu)\qquad , \qquad \vartheta_2^\mu = P^\mu ( 2q'\cdot q - q^2) - P\cdot q (2 q^{\prime\mu} - q^\mu) \ ;\\
&& \vartheta_3^\mu = \gamma^\mu q'\cdot q - \qs  q^{\prime\mu}\;\qquad , \qquad \vartheta_4^\mu = 2 \gamma^\mu P\cdot q - 2 \qs P^\mu - m_N (\gamma^\mu \qs - \qs \gamma^\mu) \ ; \\
&&\vartheta_5^\mu =  q^\mu  q' \cdot q - q^{\prime\mu} \, q^2\qquad , \qquad \!\vartheta_6^\mu = q^\mu\, \qs - \gamma^\mu \, q^2  \ ,
\label{eq:photoMampsdefined} 
\EA
and $P=(p+p')/2$. Here we have generalized $q$ to non-vanishing space-like momentum transfer, $q=(0,{\bf q}_\perp,q^-)$ while maintaining $q'=(0,{\bf 0},q^{\prime-})$. 
Therefore, $t=q^2=-{\bf q}_\perp^2$. In the case of photoproduction, only the first four amplitudes are non-vanishing whereas generally all six contribute to electroproduction.

One further defines
\BA
{\cal C}_{\alpha\beta}^{\mu}(p,p',q)& =& 
\int d^4x\,e^{i {q}\cdot x} \,\langle\, N\,;\,p'\,|\, [\,{\tilde {\cal D}}_{5\alpha}(x)\,,\, {\tilde J}_\beta^\mu(0)\,]\, |\, N \,;\,p\,\rangle \ ,
\label{eq:c5defined} 
\EA
where
\BA
&&{\cal C}^{r}_{\alpha\beta}(p,p',q) \ +\ {\cal C}^{r}_{\alpha\beta}(p,p',-q) \ =\ w_{\lambda'\lambda;\alpha\beta}^{NN;r}(p,p',q) \ .
\label{eq:crdefinedinampsrs} 
\EA
A Lorentz-invariant decomposition is
\BA
{\cal C}_{\alpha\beta}^{\mu}(p,p',q)& =& 
i{\bar u}_{\lambda'}(p')\,\gamma_5\;\sum_{i=1}^{6}\;{\vartheta}_i^\mu\;{\cal W}^i_{\alpha\beta}({\bar\nu},q^2)\;{u}_\lambda (p) \  .
\label{eq:photoCdefined} 
\EA
It is straightforward to find 
\BA
{\rm Im}\;{A}_{i\alpha\beta}({\bar\nu},q^2)  \ =\ \frac{e\pi}{F_\pi} {\cal W}_{i\alpha\beta}({\bar\nu},q^2) \ .
\label{eq:electrocdefinedinamps3} 
\EA
The crossing properties that are relevant for this paper are
\BA
{\cal W}_{1(\alpha\beta)}({\bar\nu},q^2) & =&  -{\cal W}_{1(\alpha\beta)}(-{\bar\nu},q^2) ; \\
{\cal W}_{6[\alpha\beta]}({\bar\nu},q^2) & =&  -{\cal W}_{6[\alpha\beta]}(-{\bar\nu},q^2) . 
\label{eq:electrocrossing} 
\EA
One defines
\BA
{\cal W}_{1(\alpha\beta)}({\bar\nu},q^2)   & =&  \delta_{\alpha\beta} {\cal W}^+_1({\bar\nu},q^2) \ =\ \delta_{\alpha\beta} \frac{F_\pi}{e\pi} {\rm Im}\;{A}^+_1({\bar\nu},q^2) ; \\
{\cal W}_{6[\alpha\beta]}({\bar\nu},q^2)   & =&  2 i \epsilon_{\alpha\beta\gamma} T_\gamma {\cal W}^-_6({\bar\nu},q^2) \ =\ 2 i \epsilon_{\alpha\beta\gamma} T_\gamma \frac{F_\pi}{e\pi} {\rm Im}\;{A}^-_6({\bar\nu},q^2) .
\label{eq:electrocrossingdefinew1} 
\EA
As these amplitudes occur within the nucleon (iso)spinors, it is useful to define the amplitudes with the isospinors present
\BA
{\cal A}_i({\bar\nu},q^2) &\equiv&  \xi^T\,{A}_i({\bar\nu},q^2)\,\xi \ .
\label{eq:withisos}
\EA

\section{Nucleon form factor conventions}\label{sec:NFFC}

\subsection*{Vector form factor}%\label{vectorff}

\noindent Our convention for the nucleon vector form factors is 
\BA
\langle\, N\,,\,\lambda'\,;\,p'\, |\, {\tilde J}^{\mu}_{\alpha}(0) \,|\, N\,,\,\lambda\,;\,p\, \rangle \, =\, 
{\bar u}_{\lambda'}(p')\,\Big\lbrack
\gamma^\mu\, F_{1\alpha}^V(q^2)\ +\ \frac{i\sigma^{\mu\nu}q_\nu}{2 m_N}F_{2\alpha}^V(q^2)
\,\Big\rbrack\,{u}_{\lambda}(p) \ ,
\label{eq:NvffIF}
\EA
with $q=p'-p$ and 
\BA
F_{i\alpha}^V(q^2) &\equiv & T_\alpha F^V_i(q^2) 
\label{eq:FvFi}
\EA
with $i=1,2$ and $T_\alpha=\tau_\alpha/2$, and $F^V_1(q^2) $ and $F^V_2(q^2) $  are the Dirac and Pauli form factors, respectively.
The isovector and isoscalar form factors decompose into proton and neutron form factors as:
\BA
F^V_i(q^2) \ =\  F^p_i(q^2) \ -\ F^n_i(q^2) \ \ \ , \ \ \
F^S_i(q^2) \ =\  F^p_i(q^2) \ +\ F^n_i(q^2)  \ ,
\label{eq:Nvffdecomp}
\EA
and $F^p_1(0)=1$, $F^n_1(0)=0$, $F^p_2(0)=\kappa_p$, and $F^n_2(0)=\kappa_n$,
and therefore $F^V_1(0)=F^S_1(0)=1$, $F^V_2(0)=\kappa_p-\kappa_n=\kappa_V$ and $F^S_2(0)=\kappa_p+\kappa_n=\kappa_S$. 
The electromagnetic form factors are then given by
\BA
F^{EM}_i(q^2) &=& {\textstyle \frac{1}{2}} F^S_i(q^2){\bf 1} \ +\ F^V_i(q^2)\,T_3 \ .
\label{eq:Nvffdecompem}
\EA
The $q^2$ expansion of the Dirac form factor gives
\BA
F^V_1(q^2) &=& 1\ +\ \frac{q^2}{6}\vev{r^2}^V_1 \ +\ \ldots
\label{eq:cr}
\EA
so that the isovector Dirac radius is
\BA
\vev{r^2}^V_1 \ =\ 6 \frac{dF_1^V(q^2)}{dq^2}|_{q^2=0} \ =\  6 {F_1^V}^\prime(0)\ .
\label{eq:crderiv}
\EA
The form factors can be expressed in terms of nucleon
matrix elements using the null-plane momentum eigenstates and spinor
decomposition given in Appendix~\ref{npconventions}. Noting that 
$q^+=0$ in all matrix elements that we consider, one then readily finds
\BA
\langle\, N,\uparrow\, ;\,p' |\, {\tilde J}^{+}_{\alpha}(0) \,|\, N,\uparrow\, ;\,p \rangle \, =\, 
\langle\, N;\downarrow\, ;\,p'|\, {\tilde J}^{+}_{\alpha}(0) \,|\, N;\downarrow\, ;\,p\rangle \, =\, 
2p^+ \xi^T F^V_{1\alpha}(q^2) \xi \ ;
\label{eq:NvffFF1}
\EA
\BA
\langle\, N;\uparrow\, ;\,p'|\, {\tilde J}^{+}_{\alpha}(0) \,|\, N;\downarrow\, ;\,p\rangle \, =\, 
-2p^+  \bar q_\perp \xi^T {\frac{F^V_{2\alpha}(q^2)}{2m_N}}\xi \ ;
\label{eq:NvffFF2}
\EA
\BA
\langle\, N;\downarrow\, ;\,p'|\, {\tilde J}^{+}_{\alpha}(0) \,|\, N;\uparrow\, ;\,p\rangle \, =\, 
2p^+  q_\perp \xi^T {\frac{F^V_{2\alpha}(q^2)}{2m_N}}\xi \ .
\label{eq:NvffFF2b}
\EA

\subsection*{Axial form factor}%\label{vectorA}

\noindent The nucleon axial-vector form factor is
\BA
\langle\, N\,,\,\lambda'\,;\,p'\, |\, {\tilde J}^{\mu}_{5\alpha}(0) \,|\, N\,,\,\lambda\,;\,p\, \rangle \, =\, 
{\bar u}_{\lambda'}(p')\,\Big\lbrack
\gamma^{\mu}\, G_\alpha^A(q^2)\ +\ \frac{q^\mu}{2 m_N}G_\alpha^P(q^2)
\,\Big\rbrack\,\gamma_5\,{u}_{\lambda}(p) \ .
\label{eq:NaffIF}
\EA
where 
\BA
G_\alpha^{A,P}(q^2) &=& T_\alpha G^{A,P}(q^2) \ .
\label{eq:FaFi}
\EA
The $q^2$ expansion of the axial form factor is
\BA
G^A(q^2) &=& g_A\left(\; 1\; +\; \frac{q^2}{6}\vev{r^2}^A\ +\ \ldots \;\right) 
\label{eq:ar}
\EA
so that the axial root-mean square radius is
\BA
g_A\;\vev{r^2}^A \ =\ 6 \frac{dG^A(q^2)}{dq^2}|_{q^2=0} \ =\  6 {G^A}^\prime(0)\ .
\label{eq:arderiv}
\EA
The axial form factors can be expressed in terms of nucleon
matrix elements using the null-plane momentum eigenstates and spinor
decomposition give in Appendix~\ref{npconventions}. Noting that 
$q^+=0$ in all matrix elements that we consider, one then readily finds
\BA
\langle\, N,\uparrow\, ;\,p' |\, {\tilde J}^{+}_{5\alpha}(0) \,|\, N,\uparrow\, ;\,p \rangle \, =\, 
-\langle\, N;\downarrow\, ;\,p'|\, {\tilde J}^{+}_{5\alpha}(0) \,|\, N;\downarrow\, ;\,p\rangle \, =\, 
2p^+ \xi^T G^A_{\alpha}(q^2) \xi \ ;
\label{eq:NaffFF1}
\EA
\BA
\langle\, N;\uparrow\, ;\,p'|\, {\tilde J}^{+}_{5\alpha}(0) \,|\, N;\downarrow\, ;\,p\rangle \, =\, 0 \ .
\label{eq:NaffFF2}
\EA

%\vfill\eject
\bibliographystyle{JHEP}
\bibliography{bibi}

\end{document}